\documentclass[aip,jcp,amsmath,amssymb,reprint]{revtex4-1}

\usepackage{graphicx}
\usepackage{graphicx,color}
\usepackage{dcolumn}
\usepackage{bm}

\begin{document}

\newcommand{\NewTextRevA}[1]{\textcolor{red}{#1}}
\newcommand{\NewTextRevB}[1]{\textcolor{green}{#1}}

\preprint{AIP/123-QED}

\title{Computer simulations of melts of randomly branching polymers}

\author{Angelo Rosa}
\email{anrosa@sissa.it}
\affiliation{
Sissa (Scuola Internazionale Superiore di Studi Avanzati), Via Bonomea 265, 34136 Trieste, Italy
}

\author{Ralf Everaers}
\email{ralf.everaers@ens-lyon.fr}
\affiliation{
Univ Lyon, Ens de Lyon, Univ Claude Bernard Lyon 1, CNRS, Laboratoire de Physique and Centre Blaise Pascal, F-69342 Lyon, France
}

\date{\today}

\begin{abstract}
Randomly branching polymers with {\em annealed} connectivity are model systems for ring polymers and chromosomes. In this context, the branched structure represents transient folding induced by topological constraints.
Here we present computer simulations of melts of annealed randomly branching polymers of $3\le N\le1800$ segments in $d=2$ and $d=3$ dimensions. In all cases, we perform a detailed analysis of the observed tree connectivities and spatial conformations.
Our results are in excellent agreement with an asymptotic scaling of the average tree size of  $R \sim N^{1/d}$, suggesting that the trees behave as compact, {\it territorial} fractals. The observed swelling relative to the size of ideal trees, $R\sim N^{1/4}$, demonstrates that excluded volume interactions are only partially screened in melts of annealed trees. 
Overall, our results are in good qualitative agreement with the predictions of 
Flory theory. In particular, we find that the trees swell by the combination of modified branching and path stretching.
However, the former effect is subdominant and difficult to detect in $d=3$ dimensions.
\end{abstract}

\maketitle

\section{Introduction}\label{sec:intro}
\begin{figure}
\begin{center}
\includegraphics[width=0.22\textwidth,angle=90]{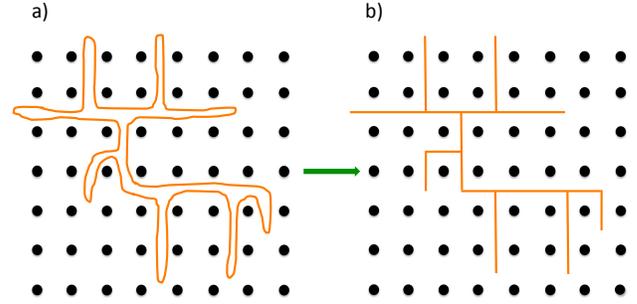}
\end{center}
\caption{
\label{fig:RubinsteinElk}
Behaviour of a two-dimensional ring (orange) in an array of impenetrable fixed obstacles (schematically represented by the black dots).
To maximize its conformational entropy, the ring adopts a transiently branched shape (a),
which can then be mapped to a lattice tree (b) with {\it annealed} connectivity.
The figure is adapted from~\cite{kapnistos2008}.
}
\end{figure}

Randomly branched polymers or trees display surprisingly rich physics.
In Statistical Mechanics, lattice trees are believed to fall into the same universality class as lattice animals~\cite{IsaacsonLubensky,SeitzKlein1981,DuarteRuskin1981} and their critical exponents are related to those of magnetic systems~\cite{ParisiSourlasPRL1981,FisherPRL1978,KurtzeFisherPRB1979,BovierFroelichGlaus1984}.
In Polymer Chemistry, the deliberate (or accidental~\cite{Read2013}) incorporation of monomers with higher functionality into the polymerisation processes modifies materials properties~\cite{RubinsteinColby,Burchard1999}. 
In this context, one has to distinguish the environmental conditions under which chains are studied from those under which they are synthesised and where their connectivity is said to be {\em quenched}.
Here we are interested in randomly branched polymers with {\em annealed} connectivity, whose structure is meant to represent the transient folding of topologically constrained ring polymers~\cite{KhokhlovNechaev85,RubinsteinPRL1986,RubinsteinPRL1994,kapnistos2008,GrosbergSoftMatter2014,RosaEveraersPRL2014,Rosa2016a}
(Fig.~\ref{fig:RubinsteinElk}) and chromosomes~\cite{grosbergEPL1993,RosaPLOS2008,Vettorel2009,MirnyRev2011}.
At the light of the recent results by Lang~\cite{LangMacromol2013} and Smrek and Grosberg~\cite{SmrekGrosbergACSMacroLett2016} who analysed the threadable fraction of the minimal area encircled by non-concatenated ring polymers in melt,
it is a non-trivial and still open question, if~\cite{CatesDeutsch} or to which extent~\cite{RubinsteinPRL1994,RubinsteinMacromolecules2016} this analogy~\cite{KhokhlovNechaev85} holds also for these systems.
However, having shown that it provides at least an excellent approximation~\cite{RosaEveraersPRL2014}, 
we now proceed to analyse  in some detail the statistical properties of melts of annealed trees.

As customary in polymer physics~\cite{DeGennesBook,DoiEdwards,KhokhlovGrosberg,RubinsteinColby}, we are primarily interested in exponents describing how expectation values for observables scale with the weight, $N$, of the trees: 
\begin{eqnarray}
\langle N_{br}(N) \rangle & \sim & N^\epsilon                           \label{eq:epsilon}\\
\langle L(N) \rangle & \sim & N^\rho                                          \label{eq:rho}\\
\langle R_g^2(N) \rangle & \sim & N^{2\nu}                              \label{eq:nu}
\end{eqnarray}
or the path distance $l$ between tree nodes:
\begin{eqnarray}
\langle R^2(l) \rangle & \sim & l^{2\nu_{\mathrm{path}}}                          \label{eq:nu_path}\\
\langle p_c(l) \rangle & \sim & l^{-\nu_{\mathrm{path}}(d+\theta_{\mathrm{path}})} \label{eq:theta_path}
\end{eqnarray}
Here, $\langle N_{br}(N)\rangle$ denotes the average branch weight; 
$\langle L(N) \rangle$ the average contour distance or length of paths on the tree; 
$\langle R_g^2(N) \rangle$ the mean-square gyration radius of the trees; 
and $\langle R^2(l) \rangle$ and $\langle p_c(l) \rangle$ the mean-square spatial distance and contact probability of nodes as a function of their contour distance, $l$.
For ideal, non-interacting trees~\cite{ZimmStockmayer49,Rosa2016a} $\rho=\epsilon=\nu_{\mathrm{path}}=1/2$, $\nu=1/4$ and $\theta_{\mathrm{path}}=0$.
For interacting systems, the only exactly known exponent is $\nu=1/2$ for self-avoiding trees in $d=3$ dimensions~\cite{ParisiSourlasPRL1981}.
Flory theory provides a simple and insightful description of a wide range of interacting tree systems~\cite{IsaacsonLubensky,DaoudJoanny1981,GutinGrosberg93,GrosbergSoftMatter2014}, 
but the results are obtained through uncontrolled approximations and rely on the cancellation of large errors 
~\cite{DeGennesBook,DesCloizeauxBook}.
For the present problem, Flory theory 
predicts that trees in a melt ought to behave as compact fractals with $\langle R_g^2(N) \rangle \sim N^{2/d}$ and that trees should swell by a combination of path swelling and modified branching.
As plausible as this prediction might be, it needs to be corroborated by more rigorous approaches.

Here we present the, to our knowledge, first computational study of the properties of melts of annealed trees in $d=2$ and $d=3$ dimensions.
The article is part of a series~\cite{Everaers2016a,Everaers2016b,Rosa2016a,Rosa2016c},
where we use a combination of computer simulations, Flory theory and scaling arguments to investigate the connectivity and conformational statistics of randomly branched polymers with excluded volume interactions.
We employ the same notation, definitions and numerical methodologies introduced in our previous work~\cite{Rosa2016a} on single self-avoiding trees in good solvent, which we briefly summarise in Sections~\ref{sec:theory} and~\ref{sec:modmethods}.
Results for trees connectivity and spatial conformations are outlined in Sec.~\ref{sec:results} and discussed in detail in Sec.~\ref{sec:Discussion}.
Finally, conclusions are sketched in Sec.~\ref{sec:concls}.

\section{Model and Background}\label{sec:theory}

We are interested in randomly branched polymers with annealed connectivity and repulsive, short-range interactions between monomers.
Secs.~\ref{sec:ModelUnits} 
and \ref{sec:observables} summarize
our choices of the employed lattice model, units and notation and the definition of observables.
Sec.~\ref{sec:Flory} reviews predictions of Flory theory for interacting trees.
Finally, Secs.~\ref{sec:BlobSize1} and~\ref{sec:BlobSize2} provide a justification for the model parameters and the sizes of simulated trees by employing the concept of ``blob size''.
All our numerical results are obtained for trees embedded in $d=2$ and $d=3$ dimensions, even though many theoretical expressions are conveniently expressed for general $d$.

\subsection{Model}\label{sec:ModelUnits}

We study lattice trees on the $2d$- and $3d$-cubic lattice. The functionality of the nodes is restricted to the values $f=1$ (a leaf or branch tip), $f=2$ (linear chain section), and $f=3$ (branch point). Connected nodes occupy adjacent lattice sites. A tree conformation, ${\cal T}\equiv ({\mathcal G},\Gamma)$, can be described by the set of node positions, $\Gamma=\{\vec r_1, \dots, \vec r_{N+1}\}$, in the embedding space and a suitable representation of its connectivity graph, ${\mathcal G}$. We employ a data structure in the form of a linked list, which retains for each node, $i$, its position, $\vec r_i$, functionality, $f_i$, and the indices $\{j_1(i),\ldots,j_{f_i}(i)\}$ of the nodes to which it is connected. 

Since our models do not include a bending energy, the lattice constant equals the Kuhn length, $l_K$, of linear paths across ideal trees. We measure energy in units of $k_BT$, length in units of the lattice constant or Kuhn length, $l_K$, and mass in units of the number of Kuhn segments. Similarly, we specify the density by the Kuhn segment number density, $\rho_K$.
We use the letters $N$ and $n$ to denote the mass of a tree or a branch, respectively.
With $N$ Kuhn segments connecting the nodes of a tree, there are $N+1$ nodes in a tree.
The symbols $L$ and $l$ are reserved for contour lengths of {\em linear} paths on the tree, while $\delta L$ and $\delta l$ denote contour distances from a fixed point, typically the tree center.
Spatial distances are denoted by the letters $R$ and $r$.
Examples are the tree gyration radius, $R_g$, spatial distances between nodes, $\vec r_{ij}$, and the spatial distances, $\vec{\delta r}_i$, of a node from the tree center of mass.

For ideal trees, nodes do not interact and their asymptotic branching probability, $\lambda$, is controlled via a chemical potential for branch points,
\begin{equation}\label{eq:HIdeal}
{\mathcal H}_{id}({\cal T}) =  \mu_{br} n_3({\mathcal G})
\end{equation}
where $n_3({\mathcal G})$ is the total number of  3-functional nodes in the tree.
All our results are obtained~\cite{RosaEveraersPRL2014} for a value of $\mu_{br} = -2.0 \, k_B T$.
Interactions between nodes are accounted for via
\begin{equation}\label{eq:HVolumeInteractions}
{\mathcal H}_{int}({\cal T}) =  v_K \sum_{i \in lattice} n_{K,i}^2
\end{equation}
where $n_{K,i}$ is the total number of Kuhn segments inside the elementary cell centered at the lattice site $i$.
In all cases we employ the same, large free energy penalty of $v_K = 4 \, k_B T$ for overlapping pairs of Kuhn segments, Eq.~(\ref{eq:HVolumeInteractions}).
The pair repulsion is so strong, that single trees are effectively self-avoiding,
while the local occupancy fluctuations in our melts is
$\langle (n_{K,i}-\langle n_{K,i} \rangle)^2 \rangle \approx 0.12-0.13$ {\it for all densities}.

\subsection{Observables}\label{sec:observables}

A tree is a branched structure free of loops. Its connectivity can be characterized in a number of ways.
Locally, nodes connecting Kuhn segments differ according to their functionality, $f$, with branch points having functionality $f\equiv3$ and branch tips $f\equiv1$.
Given $n_f$ the total number of tree nodes with functionality $f$, they satisfy the relations:
\begin{eqnarray}
n_1 &=& 2+n_3
\label{eq:n1vsn3}\\
n_2 &=& N-1-2n_3
\end{eqnarray}
For our choice of parameters, ideal trees are characterized by an asymptotic branching probability~\cite{Rosa2016a} %
\begin{equation}\label{eq:lambda}
\lambda = \lim_{N\rightarrow\infty} \langle \frac{n_3(N)}N \rangle = 0.4
\end{equation}
The large scale structure of a tree can be analyzed in terms of the ensemble of sub-trees generated by cutting bonds. 
Removal of a bond splits a tree of weight $N$ into two trees of weight $n$ and $N-1-n$. Defining $N_{br}\equiv \min(n,N-1-n)$ as the branch weight, the corresponding ensemble average
grows as a characteristic power of the tree weight, $\langle N_{br}(N) \rangle \sim N^{\epsilon}$ (Eq.~(\ref{eq:epsilon})).

Alternatively, the tree connectivity can be analyzed in terms of:
(1)
the statistics of minimal distances, $l_{ij}$, of two nodes $i,j$ along linear paths on the tree,
and its corresponding ensemble average $\langle L(N)\rangle$;
(2)
the average distance, $\langle \delta L_{\mathrm{center}}(N) \rangle$ of nodes from the central node;
(3)
the average length $\langle \delta L_{\mathrm{center}}^{\mathrm{max}}(N) \rangle$ of the longest distance from the central node.
For ensemble averages one expects (Eq.~(\ref{eq:rho})):
$\langle L(N) \rangle \sim \langle \delta L_{\mathrm{center}} \rangle \sim  \langle \delta L_{\mathrm{center}}^{\mathrm{max}}(N) \rangle \sim N^\rho$
with~\cite{MadrasJPhysA1992} $\rho=\epsilon$.
Similarly, we characterize the statistics of branches by measuring
the average branch weight, $\langle N_{br}(\delta l_{\mathrm{root}}^{\mathrm{max}}) \rangle$,
as a function of the {\it longest} contour distance of nodes from the branch root, $\delta l_{\mathrm{root}}^{\mathrm{max}}(N_{br})$.
Finally, we consider the average weight of the ``core'' of tree, $\langle N_{center}(\delta l_{\mathrm{center}}) \rangle$,
made of segments whose distance from the central node does not exceed $\delta l_{\mathrm{center}}$.

The overall spatial extension of the tree is best described through the mean-square gyration radius $\langle R_g^2 \rangle$
which is the average square distance of a node from the tree center of mass.
Asymptotically, it is expected to scale as $\langle R_g^2 \rangle  \sim N^{2\nu}$ (Eq.~(\ref{eq:nu})).

The spatial conformations of linear paths of length $l$ on trees of total mass $N$ can be characterized using standard observables for linear polymers,
namely:
(1)
the mean-square end-to-end distance $\langle R^2(l,N) \rangle$;
(2)
for a given contact distance $r_c$, the corresponding end-to-end closure probability $\langle p_c(l,N) \rangle$.
They are expected to scale, respectively, as (Eqs.~(\ref{eq:nu_path}) and~(\ref{eq:theta_path})):
$\langle R^2(l) \rangle  \sim l^{2\nu_{\mathrm{path}}}$ and
$\langle p_c(l) \rangle \sim l^{-\nu_{\mathrm{path}} (d+\theta_{\mathrm{path}})}$
and to be asymptotically independent of tree weight. 
By construction~\cite{Rosa2016a}, $\nu = \nu_{\mathrm{path}} \, \rho$.

\subsection{Flory theory}\label{sec:Flory}

%

Flory theories~\cite{FloryChemBook} are formulated as a balance of an entropic elastic term and an interaction energy:
\begin{equation}\label{eq:fFlory}
{\mathcal F} = {\mathcal F_{el}}+{\mathcal F_{inter}} \, .
\end{equation}
The central element of the Flory theory of interacting trees 
is the elastic free energy of ideal annealed trees~\cite{GutinGrosberg93,GrosbergNechaev2015}, 
\begin{equation}\label{eq:fGutin}
\frac{{\mathcal F}_{el}}{k_B T} \sim \frac{R^2}{l_K L} +  \frac{L^2}{N\, l_K^2}  \ ,
\end{equation}
Ensembles of quenched trees are characterised by fixed values of $L\sim l_K N^\rho$ and the Flory energy needs to be minimized over $R$. Linear chains represent a special case, where $L=l_K N$. In annealed trees, interactions can modify the branching statistics as well as the spatial conformations. With larger contour distances leading to larger spatial distances between repelling monomers, the Flory energy needs to be simultaneously minimized over both variables, $R$ and $L$. Optimising $L$ for a {\em given} size, $R \sim N^\nu$, yields
\begin{eqnarray}
\label{eq:rho_of_nu}
\rho &=&\frac{1+2\nu}3
\end{eqnarray}
and
\begin{eqnarray}
\label{eq:nupath_of_nu}
\nu_{\mathrm{path}}&=& \frac{3\nu}{1+2\nu}\ .
\end{eqnarray}
Thus independently of the physical origin of the effect, swollen trees with $\nu>\nu^{\mathrm{ideal}}\equiv1/4$ are predicted to display both, modified connectivities with $\rho>\rho^{\mathrm{ideal}}\equiv1/2$ and path swelling with $\nu_{\mathrm{path}}>\nu^{\mathrm{ideal}}_{\mathrm{path}}\equiv1/2$.

In melts, volume interactions are screened~\cite{IsaacsonLubensky,GrosbergSoftMatter2014} and dominated by high-order collisions of dense systems.
Inspecting all terms of order $p$ in a (standard) virial-type expansion of the interaction energy term in Eq.~(\ref{eq:fFlory})~\cite{DaoudJoanny1981}: 
\begin{equation}\label{eq:IntTermMelt}
{\mathcal F_{inter}(N,R)} \sim \left( \frac{N}{R^d} \right)^{p-1} \, ,
\end{equation}
shows that interactions are estimated to be irrelevant, if $\nu d > 1$. Even without swelling, this is the case in $d > 4$ dimensions, where $\nu^{\mathrm{ideal}} d = d/4 > 1$ suggests ideal tree behavior. In $d \le 4$ dimensions,  for $1/4<\nu\le1/d$ the series is dominated by the $p \rightarrow \infty$ limit.
Minimising the sum of Eqs.~(\ref{eq:fGutin}) and~(\ref{eq:IntTermMelt}) in this limit with respect to $L$ and $R$  yields:
\begin{eqnarray}
\label{eq:nu_of_d_melt}
\nu &=&\frac{1}{d}\\
\label{eq:nupath_of_d_melt}
\nu_{\mathrm{path}}&=& \frac{3}{d+2}\\
\label{eq:rho_of_d_melt}
\rho&=& \frac{d+2}{3d}
\end{eqnarray}
Eqs.~(\ref{eq:nu_of_d_melt}) to~(\ref{eq:rho_of_d_melt}) predict that in the melt state annealed trees are compact fractals.
Interestingly, $\nu_{\mathrm{path}}$ has the same value as the critical exponent $\nu$ for linear self-avoiding walks,
suggesting a deeper analogy between the two problems~\cite{GrosbergSoftMatter2014}.

\subsection{Blob size}\label{sec:BlobSize1}
Assuming that entropic effects are too small to induce density inhomogeneities,
we can estimate the asymptotic behaviour from the assumption~\cite{RubinsteinPRL1994} that the asymptotic segment self density at the tree center of mass converges to the melt segment density $\rho_K$.
More precisely~\cite{RosaEveraersPRL2014}, $\phi\equiv\left( \rho_{tree} \sqrt{(2\pi)^d \det(S)} \right)^{-1}\sim N/R^d/\rho_K \rightarrow1$,
where $S$ is the gyration or shape tensor, $S_{\alpha\beta}=\frac1{N+1}\sum_{i=1}^{N+1} (\vec r_{i\alpha}-\vec r_{CM, \alpha})(\vec r_{i\beta}-\vec r_{CM, \beta})$
and $\vec r_{CM}$ is the spatial position of the tree centre of mass.
Neglecting asphericity, 
\begin{equation}
\langle R_g^2(N)\rangle_{melt}\rightarrow \frac d{2\pi} \left(\frac N{\rho_K} \right)^{2/d}\ .
\end{equation}
while our simulation results (Fig.~\ref{fig:Rg2AspectRatios}B) suggest in $d=3$ with 
$\langle R_g^2(N)\rangle_{melt}\rightarrow 0.59 \left(\frac N{\rho_K} \right)^{2/3} > \frac 3{2\pi} \left(\frac N{\rho_K}\right)^{2/3} \approx 0.48 \left(\frac N\rho \right)^{2/3}$ a slightly larger value.
By using the latter quantity, the blob size~\cite{BlobSizeNote}
\begin{equation}\label{eq:blob3d}
g \approx 0.18 \left(\rho_K l_K^3 \right)^4/\lambda^3
\end{equation}
where we expect the crossover from the ideal to the asymptotic regime to occur is implicitly defined via 
$\langle R^2(g) \rangle_{ideal} = \langle R^2(g) \rangle_{melt}$ where~\cite{ZimmStockmayer49,DaoudJoanny1981}
$\langle R^2(g) \rangle_{ideal} = \frac14 \sqrt{\frac\pi\lambda}l_K^2 g^{1/2}$.
Corresponding arguments in $d=2$ dimensions yield
$\langle R_g^2(N) \rangle_{melt}\rightarrow 0.37 \frac N{\rho_K} > \frac 1{\pi} \frac N{\rho_K} \approx 0.32 \frac N{\rho_K}$ with a blob size
\begin{equation}\label{eq:blob2d}
g \approx 1.43 \left(\rho_K l_K^2 \right)^2/\lambda \, .
\end{equation}

\subsection{Choice of simulated tree sizes and segment densities}\label{sec:BlobSize2}

Our original simulation of tree melts in $d=3$ dimensions reported in Ref.~\cite{RosaEveraersPRL2014} were carried out at a segment density of  $\rho_K = 5 l_K^{-3}$,
which was imposed by the mapping to the corresponding ring polymer problem.
In this case, the estimated blob size of $g=1758$ equals the size, $N_{\mathrm{max}}=1800$, of the largest trees we were able to simulate.
Below we report results for measured gyration radii {\em etc.} for these systems, but we exclude them from the estimation of asymptotic exponents.

Instead, we rely in $d=3$ dimensions on data obtained for a smaller density of $\rho_K = 2 l_K^{-3}$.
With $g=45 \ll N_{\mathrm{max}}=900$ we safely expect to have reached the asymptotic regime.
Similarly, in $d=2$ dimensions for $\rho_K = 2 l_K^{-2}$, we get $g=14 \ll N_{\mathrm{max}}=900$.

\section{Methods}\label{sec:modmethods}

To simulate melts of annealed lattice trees we have used a variant of the ``amoeba'' Monte Carlo algorithm by Seitz and Klein~\cite{SeitzKlein1981} (Sec.~\ref{sec:AmoebaAlgo}).
The quantitative analysis of tree connectivities, tree spatial conformations and the estimation of critical exponents defined in Eqs.~(\ref{eq:epsilon})-(\ref{eq:nu_path})
has been carried out by applying the ``burning algorithm''~\cite{StanleyJPhysA1984,StaufferAharonyBook} (Sec.~\ref{sec:TreesAnalysis}) and ordinary fitting procedures (Sec.~\ref{sec:ExtractExps}).
Tabulated values and other details on the derivation of critical exponents for single-tree statistics are also reported in the Supplemental Material~\cite{SupplMatNote}.

\subsection{Monte Carlo simulations of annealed tree melts}\label{sec:AmoebaAlgo}

\begin{figure}
\begin{center}
\includegraphics[width=0.5\textwidth]{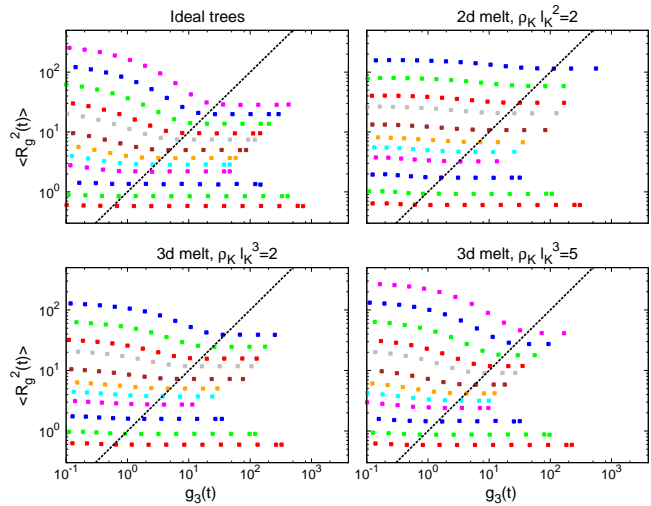}
\end{center}
\caption{
\label{fig:MC_Equilibration}
Parametric plots of the MC-time ($t$) evolution
of the ensemble-average square gyration radius, $\langle R_g^2(t) \rangle$,
{\it vs.} the mean-square displacement of the tree center of mass, $g_3(t)$.
Non-equilibrated (resp., equilibrated) values of the plots correspond
to regions above (resp., below) the black solid line $y=x$.
Different colors correspond to different tree masses ranging from
(bottom symbols) $N=3$ to (top symbols) $N=900$ (melts for $\rho_K l_K^3=2$) and $N=1800$ (ideal trees and melts for $\rho_K l_K^3=5$)}.
\end{figure}

Amoeba trial moves simultaneously modify the tree connectivity, ${\mathcal G}$, and the tree conformation, $\Gamma$.
They are constructed by randomly cutting a leaf (or node with functionality $f=1$) from the tree and placing it on a randomly chosen site adjacent to a randomly chosen node with functionality $f<3$, to which the leave is then connected.
Trial moves, ${\cal T}_i \rightarrow {\cal T}_f$, are accepted with probability:
\begin{equation}\label{eq:accRatioIdeal}
\mbox{acc}_{i \rightarrow f} =
\min \left\{ 1, \frac{n_1(i)}{n_1(f)}
e^{-\beta \left( {\mathcal H}({\cal T}_f) - {\mathcal H}({\cal T}_i)\right)} \right\}
\end{equation}
where $n_1(i/f)$ is the total number of 1-functional nodes in the initial/final state
and ${\mathcal H}({\cal T}) = {\mathcal H}_{id}({\cal T}) + {\mathcal H}_{int}({\cal T})$, Eqs.~(\ref{eq:HIdeal}) and~(\ref{eq:HVolumeInteractions}).
It should be noted, that our version of the amoeba algorithm is slightly modified with respect to the original one of Ref.~\cite{SeitzKlein1981} as we impose~\cite{RosaEveraersPRL2014} node functionalities $f\leq 3$.
Similar algorithms displacing entire branches are more efficient for single trees~\cite{MadrasJPhysA1992},
but are likely to encounter difficulties when generalized to the dense systems we are mostly interested in.
In contrast, the small non-local mass transport of the amoeba algorithm is not obstructed by the volume interactions,
since it falls into the range of the natural occupancy fluctuations in our tree melts.

Our simulations start from linearly connected random walks as initial states.
The total computational effort for equilibrating the systems as a function of the corresponding systems sizes is summarised in Table~SI~\cite{SupplMatNote}. 
As illustrated by Fig.~\ref{fig:MC_Equilibration}, the tree gyration radii equilibrate over a time scale during which the tree centers of mass diffuse over the corresponding distance. 
Quite curiously, the performance of the ``amoeba'' algorithm as a function of the Monte Carlo time steps is non-monotonic in the system density $\rho_K$, see Fig.~S1~\cite{SupplMatNote}.
In fact, while $3d$ self-avoiding trees ($\rho_K l_K^3 \rightarrow 0$, studied in our previous work~\cite{Rosa2016a}) and $3d$ melts for $\rho_K l_K^3 = 5$
take roughly the same time to reach equilibrium,
$2d$ and $3d$ tree melts at density $\rho_K l_K^d = 2$ require simulations which are $\approx 10$ times longer.
In particular,
this is the reason why we have only been able to reach tree weights up to $N_{\mathrm{max}}=900$ compared to $N_{\mathrm{max}}=1800$ in Ref.~\cite{RosaEveraersPRL2014}.
At the moment we have no explanation accounting for this, apparently counterintuitive, behavior.

\subsection{Analysis of tree connectivity}\label{sec:TreesAnalysis}


As in our previous work~\cite{Rosa2016a},
we have analysed tree connectivities using a variant of the ``burning'' algorithm for percolation clusters~\cite{StanleyJPhysA1984,StaufferAharonyBook}.
The algorithm is very simple, and consists of two parts. 
In the initial inward (or burning) pass branch tips are iteratively ``burned'' until the tree center is reached. In the subsequent outward pass one advances from the center towards the periphery. The inward pass provides information about the mass and shape of branches. The outward pass allows to reconstruct the distance of nodes from the tree center.
By employing a data structure in the form of a linked list, which retains for each node, $i$, its position, $\vec r_i$, functionality, $f_i$, and the indices $\{j_1(i),\ldots,j_{f_i}(i)\}$ of the nodes to which it is connected,
each step of the burning algorithm consists in removing from the list all sites with functionality $=1$ (tips) and updating the functionalities and the indices of the remaining nodes accordingly.
The algorithm stops when only one node remains in the list.
In order to find the minimal path length $l$ between any pair of tree nodes $i$ and $j$,
we have modified the algorithm by requiring that sites $i$ and $j$ are not removed from the list.
Accordingly, the algorithm stops when nodes $i$ and $j$ are the only tips left of the ``burned'' tree.
By using the remaining linked list it is then trivial to find the corresponding path length $l$.

\subsection{Extracting exponents from data for finite-size trees}\label{sec:ExtractExps}

In order to get reliable estimates of critical exponents ``$\rho, \epsilon, \nu_{\mathrm{path}}, \nu$'' in the large-$N$ limit and of the corresponding errors,
we stay close to the procedure developed by Janse van Rensburg and Madras~\cite{MadrasJPhysA1992} and employed in our previous work~\cite{Rosa2016a},
and combine the results obtained from fitting the $N$-dependent data to two functional forms.
For the specific example of the exponent $\nu$, they are given by the following expressions:
\begin{enumerate}
\item
A simple power-law behavior with $2$ ($a$, $\nu$) fit parameters:
\begin{equation}\label{eq:FitRg22}
\log \langle R_g^2(N) \rangle = a + 2 \nu \log N \, ,
\end{equation}
and
\item
A power-law behavior with a correction-to-scaling term ($\sim N^{-\Delta_0}$) with $4$ ($a$, $b$, $\Delta$, $\nu$) fit parameters:
\begin{eqnarray}\label{eq:FitRg21lin2}
\lefteqn{\log \langle R_g^2(N) \rangle  =} \nonumber\\
&& a + b N^{-\Delta_0} + 2 \nu \log N   - b  (\Delta-\Delta_0) N^{-\Delta_0 } \log N \, . \nonumber\\
\end{eqnarray}
Here, we have carried out a one-dimensional search for the value of $\Delta_0$ for which the fit yields a vanishing $N^{-\Delta_0 } \log N$ term.
\end{enumerate}
For the other exponents, we have employed analogous expressions.
Eq.~(\ref{eq:FitRg22}) has been used on data with $N\geq230$ and $N\geq450$ for $\rho_K l_K^d=2$ ($d=2,3$) and $\rho_K l_K^3=5$, respectively.
Eq.~(\ref{eq:FitRg21lin2}) has been employed on the whole range with $N\geq10$.
Best fits are obtained by minimizing~\cite{NumericalRecipes}
$\chi^2 = \sum_{i=1}^{D} \left[ \frac{\log \langle R_g^2(N_i)\rangle_{observed} - \log \langle R_g^2(N_i)\rangle_{model}}{\delta \log \langle R_g^2(N_i) \rangle} \right]^2$, where $D$ is the number of data points used in the fit procedure.
Quality of the fit is estimated by the normalized ${\tilde \chi}^2 \equiv \frac{\chi^2}{D-f}$,
where $f$ is the number of fit parameters.
When ${\tilde \chi}^2 \approx 1$ the fit is deemed to be reliable~\cite{NumericalRecipes}.
The corresponding $\mathcal Q(D-f, \chi^2)$-values provide a quantitative indicator for the likelihood that $\chi^2$ should exceed the observed value, if the model were correct~\cite{NumericalRecipes}.
All fit results are reported together with the corresponding errors, ${\tilde \chi}^2$ and $\mathcal Q$ values. 
Final estimates of critical exponents are calculated as averages of all independent measurements.
Corresponding uncertainties are given in the form
$\pm$(statistical error)$\pm$(systematic error),
where the ``statistical error'' is the largest value obtained from the different fits~\cite{MadrasJPhysA1992} while the ``systematic error'' is the spread between the single estimates, respectively.
In those cases where Eq.~(\ref{eq:FitRg21lin2}) fails producing trustable results we have retained only the $2$-parameter fit, Eq.~(\ref{eq:FitRg22}),
and a separate analysis of uncertainties was required, see the caption of Table~SVII~\cite{SupplMatNote} 
for details.
Error bars reported in Table~\ref{tab:ExpSummary} are given by
$\sqrt{(\mbox{statistical error})^2 + (\mbox{systematic error})^2}$.

\section{Results}\label{sec:results}

In the following sections,
we discuss the structure of trees in $2d$ and $3d$ melts by considering the scaling behaviors of the observables defined in Sec.~\ref{sec:observables}
and their corresponding critical exponents.
Similarly to the general outline of our previous work~\cite{Rosa2016a} on single self-avoiding trees in good solvent,
it is particularly instructive to compare the properties of interacting trees to the ones for ideal trees.
For this reason, but also for keeping this work autonomous and self-consistent in its own,
all figures and tables of this work, including the supplementary ones~\cite{SupplMatNote}, contain the same data for ideal trees originally reported in~\cite{Rosa2016a}.

\subsection{Branching statistics for trees with annealed connectivity}\label{sec:BranchStat}

\begin{figure}
\begin{center}
\includegraphics[width=0.5\textwidth]{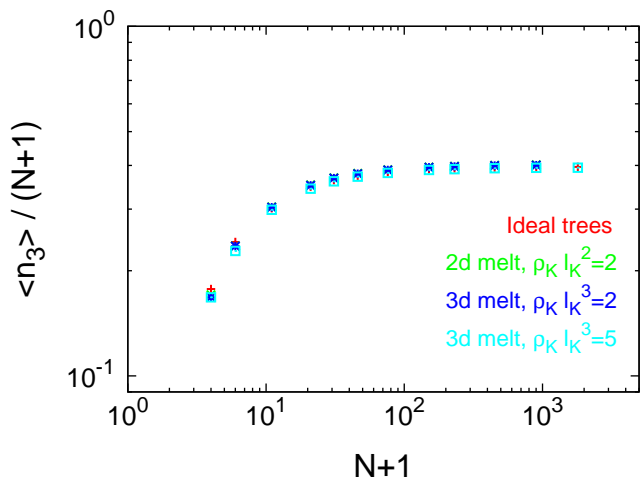}
\end{center}
\caption{
\label{fig:BranchNumber}
Branching statistics.
Average fraction of $3$-functional nodes, $\frac{\langle n_3 \rangle}{N+1}$, as a function of the total number of tree nodes, $N+1$.
The asymptotic branching probability~\cite{Rosa2016a} $\lambda$ is $=0.4$.
}
\end{figure}

Our results for the average number of branch points, $\langle n_3(N) \rangle$, as a function of $N$ are listed in Tables~SII-III~\cite{SupplMatNote}. 
Figure~\ref{fig:BranchNumber} shows that the ratios of 3-functional nodes, $\langle n_3(N) \rangle /N$,
reach their asymptotic value already for moderate tree weights.
Interestingly,
our results for ideal trees as well as interacting trees perfectly agree to each other with asymptotic branching probability~\cite{Rosa2016a} 
$\lambda = \lim_{N\rightarrow\infty} \langle n_3(N) \rangle /N = 0.4$.
In fact, due to the multiple occupation of lattice sites in the melt, the branching probabilities remain virtually unchanged in spite of the interactions.
Corresponding distributions $p(n_3)$ are well described by Gaussian statistics
with corresponding variances increasing linearly with $N$, see Fig.~S2~\cite{SupplMatNote}. 

\subsection{Path length statistics for trees}\label{sec:ppStat}

\begin{figure}
\begin{center}
\includegraphics[width=0.5\textwidth]{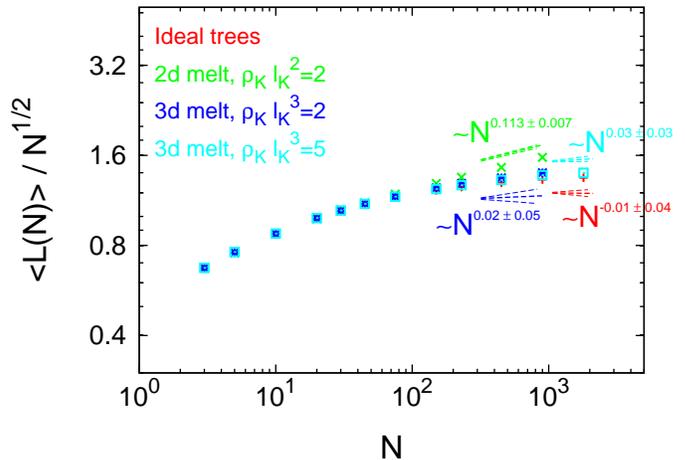}
\end{center}
\caption{
\label{fig:PathLengthStat_L}
Path length statistics.
Mean contour distance between pairs of nodes, $\langle L(N) \rangle$.
Straight lines correspond to the large-$N$ behaviour $\langle L(N) \rangle \sim N^{\rho}$ with critical exponents $\rho$
given by the best estimates summarised in Table~\ref{tab:ExpSummary}.
}
\end{figure}

Our results for 
(A)
the mean contour distance between pairs of nodes, $\langle L(N) \rangle$,
(B)
the mean contour distance of nodes from the central node, $\langle \delta L_{\mathrm{center}}(N) \rangle$, and
(C)
the mean {\it longest} contour distance of nodes from the central node, $\langle \delta L_{\mathrm{center}}^{\mathrm{max}}(N) \rangle$
are summarized in Tables~SII-III~\cite{SupplMatNote} 
and plotted in Fig.~\ref{fig:PathLengthStat_L} and Fig.~S3~\cite{SupplMatNote}. 
As discussed in Sec.~\ref{sec:observables}, the three quantities are expected to scale with the total tree weight $N$ as
$\langle \delta L_{\mathrm{center}}(N) \rangle \sim \langle \delta L_{\mathrm{center}}^{\mathrm{max}}(N) \rangle \sim \langle L(N) \rangle \sim N^{\rho}$.
Extracted single values for $\rho$'s including more details on their statistical significance ($\chi^2$ and $\mathcal Q$-values) are summarized in Table~SIV~\cite{SupplMatNote}. 
Our final best estimates for $\rho$'s (straight lines in Fig.~\ref{fig:PathLengthStat_L} and Fig.~S3~\cite{SupplMatNote}, 
and Tables~\ref{tab:ExpSummary} and~SIV~\cite{SupplMatNote}) 
are obtained by combining the corresponding results for $\langle \delta L_{\mathrm{center}}(N) \rangle$, $\langle \delta L_{\mathrm{center}}^{\mathrm{max}}(N) \rangle$ and $\langle L(N) \rangle$.

\subsection{Path lengths vs. weights for branches}\label{sec:ppStatBranches}

\begin{figure*}
\includegraphics[width=0.8\textwidth]{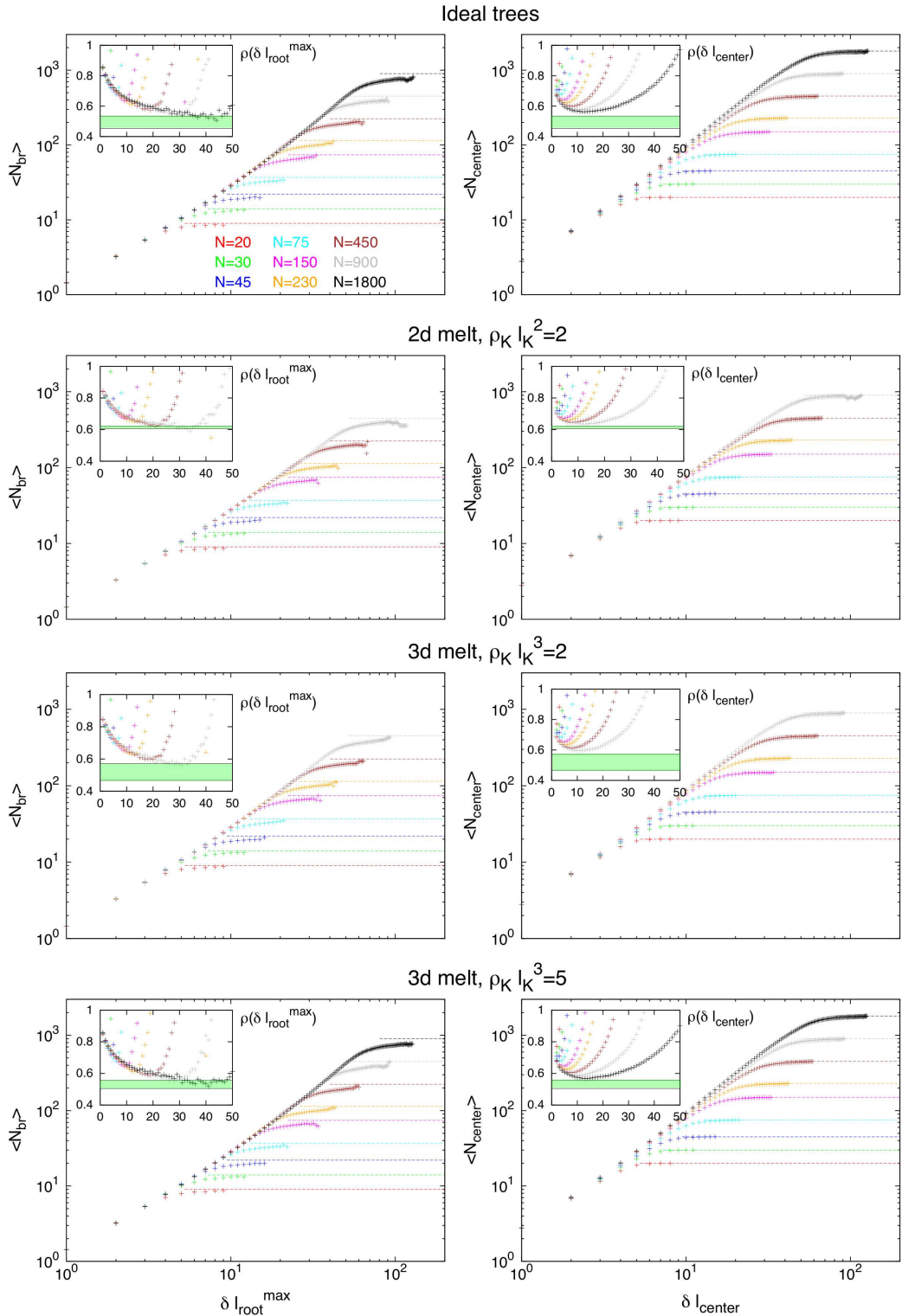}
\caption{
\label{fig:avMass_VS_lbranch}
Path lengths vs. weights of branches.
Data for trees of weight from $N=20$ to $N=1800$.
(Left)
Average branch weight, $\langle N_{br}(\delta l_{\mathrm{root}}^{\mathrm{max}})\rangle$, as a function of the longest contour distance to the branch root, $\delta l_{\mathrm{root}}^{\mathrm{max}}$.
For large $\delta l_{\mathrm{root}}^{\mathrm{max}}$, curves saturate to the corresponding maximal branch weight $(N-1)/2$ (dashed horizontal lines).
(Right)
Average branch weight, $\langle N_{center}(\delta l_{\mathrm{center}})\rangle$, composed of segments whose distance from the central node does not exceed $\delta L_{\mathrm{center}}$.
For large $\delta l_{\mathrm{center}}$, curves saturate to the corresponding total tree weight, $N$ (dashed horizontal lines).
Insets:
Corresponding differential fractal exponent $\rho(\delta l_{\mathrm{root}}^{\mathrm{max}})$ and $\rho(\delta l_{\mathrm{center}})$.
Shaded regions show the range of $\rho$ values summarized in Table~\ref{tab:ExpSummary}.
}
\end{figure*}

The relation between branch weight and path length can also be explored on the level of branches.
We have analyzed the scaling behavior of:
(1) the average branch weight, $\langle N_{br}(\delta l_{\mathrm{root}}^{\mathrm{max}})\rangle$, as a function of the longest contour distance to the branch root, $\delta l_{\mathrm{root}}^{\mathrm{max}}$, and
(2) the average branch (or tree core) weight, $\langle N_{\mathrm{center}}(\delta l_{\mathrm{center}})\rangle$, inside a contour distance $\delta l_{\mathrm{center}}$ from the central node of the tree.
Corresponding results are shown in Fig.~\ref{fig:avMass_VS_lbranch}.
Both data sets show universal behavior at intermediate $\delta l_{\mathrm{root}}^{\mathrm{max}}$ and $\delta l_{\mathrm{center}}$,
and saturate to the corresponding expected limiting values $=\frac{N-1}{2}$ and $=N$.
For large $\delta l_{\mathrm{root}}^{\mathrm{max}}$ (respectively, $\delta l_{\mathrm{center}}$) the relation $\langle N_{br}(\delta l_{\mathrm{root}}^{\mathrm{max}}) \rangle \sim {\delta l_{\mathrm{root}}^{\mathrm{max}}}^{1/\rho}$
(resp., $\langle N_{\mathrm{center}} (\delta l_{\mathrm{center}}) \rangle \sim {\delta l_{\mathrm{center}}}^{1/\rho}$) is expected to hold.
For $N_{br}$ (and with an analogous expression for $N_{center})$, we have estimated $\rho$ as
$\rho(\delta l_{\mathrm{root}}^{\mathrm{max}}) =
\left(
\frac
{\log \, \langle N_{br}(\delta l_{\mathrm{root}}^{\mathrm{max}}+1) \rangle / \langle N_{br}(\delta l_{\mathrm{root}}^{\mathrm{max}}) \rangle}
{\log \, (\delta l_{\mathrm{root}}^{\mathrm{max}}+1) / \delta l_{\mathrm{root}}^{\mathrm{max}}} \right)^{-1} $.
Numerical results are reported in the corresponding insets of Fig.~\ref{fig:avMass_VS_lbranch},
the large-scale behaviour agreeing well with the best estimates for $\rho$'s (horizontal lines) summarized in Table~\ref{tab:ExpSummary}.

\subsection{Branch weight statistics}\label{sec:BWeightStat}

\begin{figure}
\begin{center}
\includegraphics[width=0.5\textwidth]{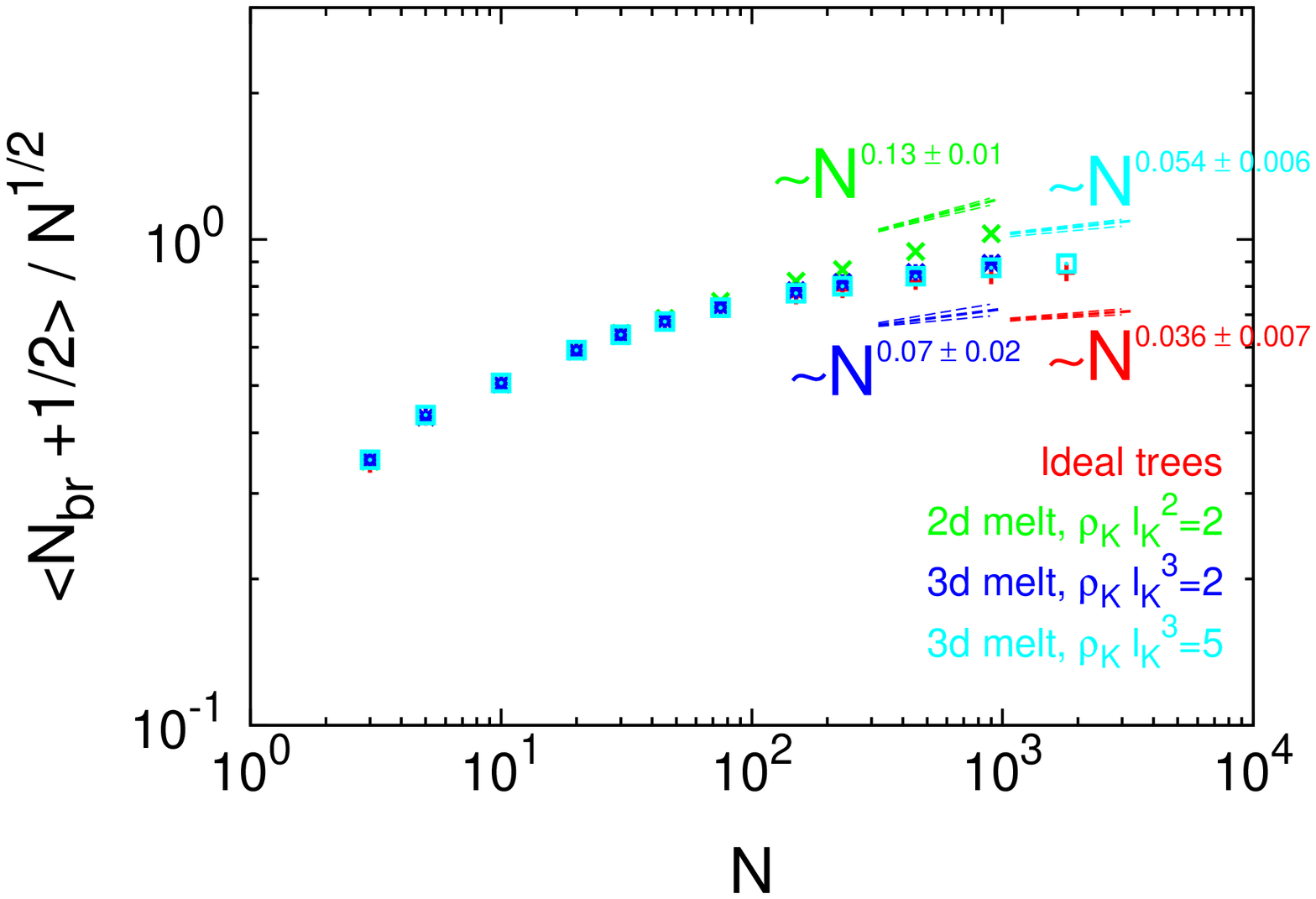}
\end{center}
\caption{
\label{fig:nBranch}
Branch weight statistics.
Average branch weight, $\langle N_{br} (N) \rangle$ as a function of the total tree mass, $N$.
Straight lines correspond to the large-$N$ behaviour $\langle N_{br} (N) \rangle \sim N^{\epsilon}$ with critical exponents $\epsilon$
given by the best estimates summarised in Table~\ref{tab:ExpSummary}.
}
\end{figure}

The scaling behavior of the average branch weight, $\langle N_{br} (N) \rangle \sim N^{\epsilon}$, defines the critical exponent $\epsilon$.
Single values of $\langle N_{br} (N) \rangle$ for each $N$ (see Tables~SII-III~\cite{SupplMatNote}) 
are plotted in Fig.~\ref{fig:nBranch},
where the straight lines have slopes corresponding to our best estimates for $\epsilon$'s (Table~\ref{tab:ExpSummary}), see also Table~SIV~\cite{SupplMatNote} 
for details.
We notice, in particular, that the scaling relation $\rho = \epsilon$ holds within error bars.

\subsection{Conformational statistics of linear paths}\label{sec:ConfStatPaths}

\begin{figure*}
\includegraphics[width=\textwidth]{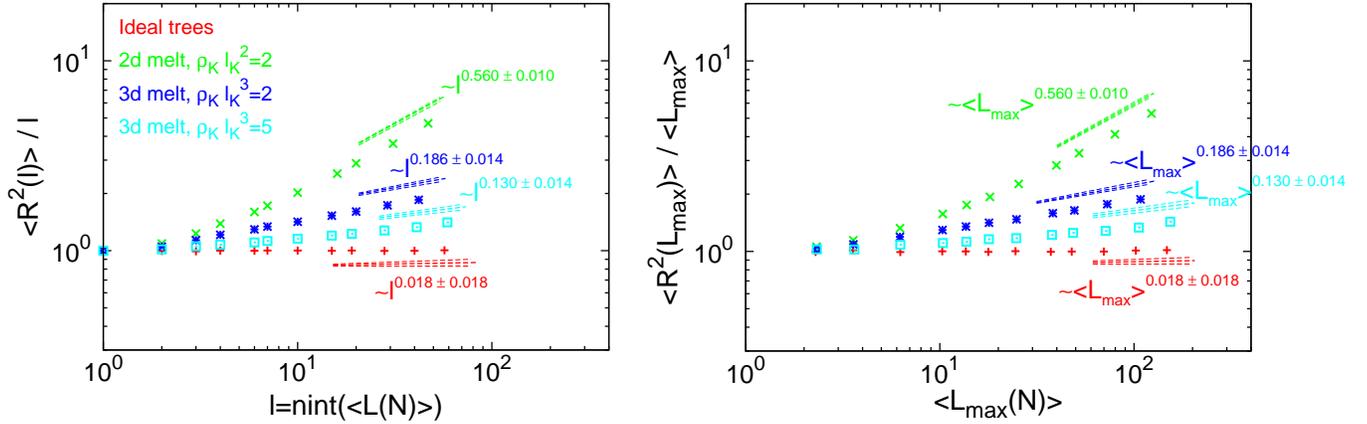}
\caption{
\label{fig:R2MaxLmax}
Conformational statistics of linear paths.
(Left)
Mean-square end-to-end distance, $\langle R^2(\ell={\tt nint}(\langle L \rangle)) \rangle$, of paths of length $\ell = {\tt nint} (\langle L(N) \rangle) \equiv$ closest-integer-to $\langle L(N) \rangle$.
(Right)
Mean-square end-to-end distance, $\langle R^2(L_{\mathrm{max}}) \rangle$, of the longest paths.
Straight lines correspond to the large-$\ell$ or large-$\langle L_{\mathrm{max}}(N) \rangle$ behaviours:
$\langle R^2(\ell) \rangle \sim \ell^{2 \nu_{\mathrm{path}}}$ or
$\langle R^2(L_{\mathrm{max}}) \rangle \sim \langle L_{\mathrm{max}}(N) \rangle^{2 \nu_{\mathrm{path}}}$.
Critical exponents $\nu_{\mathrm{path}}$ are given by the best estimates reported in Table~\ref{tab:ExpSummary}. 
}
\end{figure*}

\begin{figure*}
\begin{center}
\includegraphics[width=0.8\textwidth]{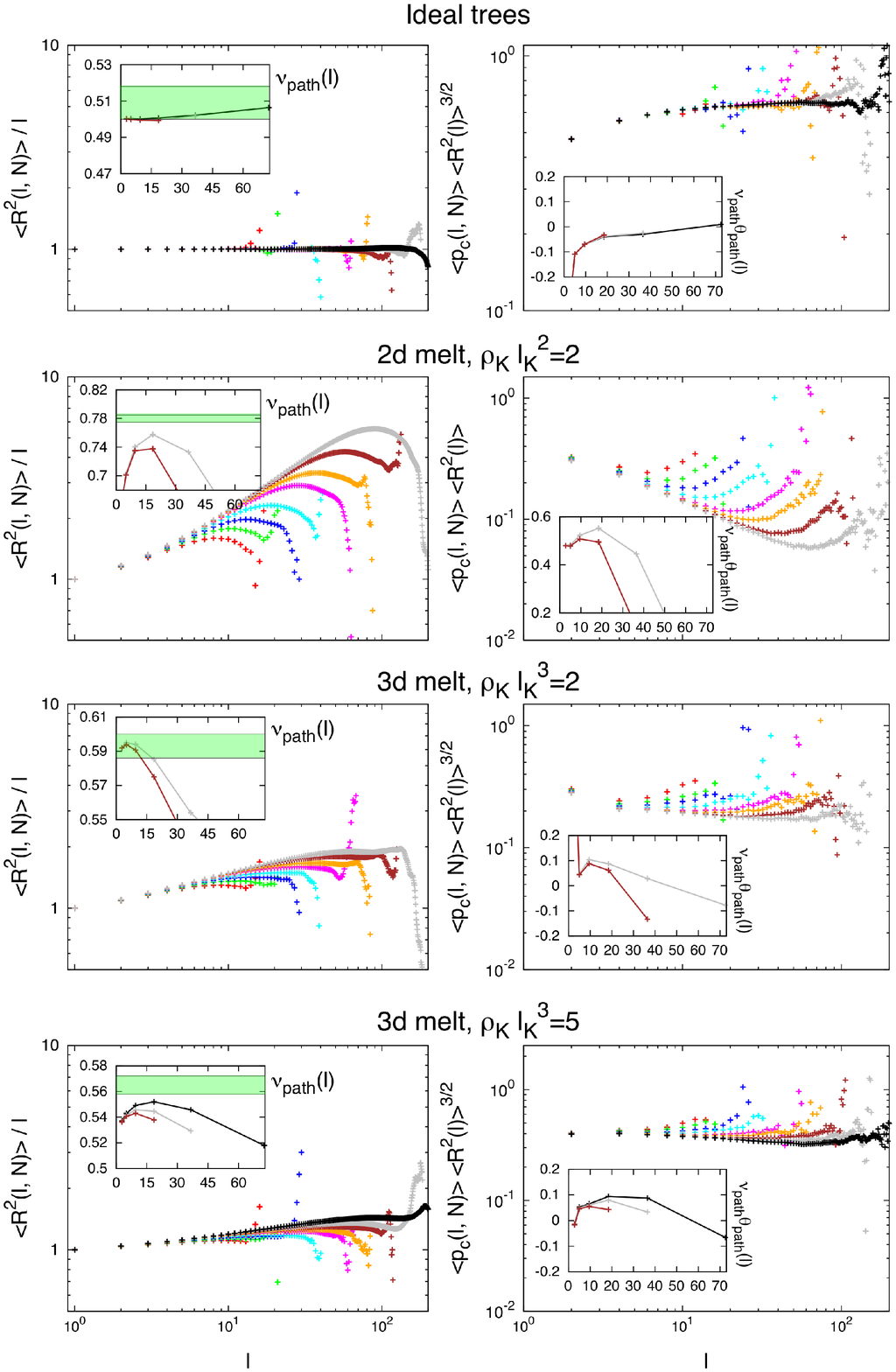}
\end{center}
\caption{
\label{fig:PathR2_ContactFreq}
Conformational statistics of linear paths.
(L.h. column)
Mean-square end-to-end distance, $\langle R^2(l, N) \rangle$, of linear paths of length $l$.
Insets:
Differential fractal exponent,
$\nu_{\mathrm{path}}(l) = \frac{1}{2}\frac{\log \, \langle R^2(l+1, N) \rangle / \langle R^2(l, N) \rangle}{\log \, (l+1)/l}$ for chain lengths $N\geq450$.
Shaded regions show the range of $\nu_{\mathrm{path}}$ values summarized in Table~\ref{tab:ExpSummary}.
(R.h. column)
Mean closure probabilities, $\langle p_c(l, N) \rangle$, between ends of linear paths of length $l$ normalised to the mean-field expectation value $\langle R^2(l, N) \rangle^{-3/2}$.
Insets:
Differential fractal exponent $\nu_{\mathrm{path}} \theta_{\mathrm{path}}(l)$, see Eq.~(\ref{eq:theta_path}), defined analogously to $\nu_{\mathrm{path}}(l)$ for chain lengths $N\geq450$.
Plots in the insets have been obtained by averaging corresponding quantities over log-spaced intervals.
Color code is as in Fig.~\ref{fig:avMass_VS_lbranch}.
}
\end{figure*}

In order to extract the critical exponent $\nu_{\mathrm{path}}$ which defines the scaling behavior $\langle R^2(l, N) \rangle \sim l^{2\nu_{\mathrm{path}}}$,
we have selected paths of length close to the average length ($l=\langle L(N) \rangle$) and to the trees maximal length ($l=L_{\mathrm{max}}(N)$)
and calculated corresponding mean-square end-to-end distances $\langle R^2(\langle L \rangle) \rangle \sim \langle L(N) \rangle^{2\nu_{\mathrm{path}}}$ and $\langle R^2(L_{\mathrm{max}}) \rangle \sim \langle L_{\mathrm{max}}(N) \rangle^{2\nu_{\mathrm{path}}}$
(see Fig.~\ref{fig:R2MaxLmax}
and corresponding tabulated values in Tables~SV-VI~\cite{SupplMatNote}. 
Combination of the two (Table~SVII~\cite{SupplMatNote}) 
led to our best estimates for $\nu_{\mathrm{path}}$ in the different ensembles summarized in Table~\ref{tab:ExpSummary}.
Not surprisingly, these values agree well with the differential exponents
$\nu_{\mathrm{path}}(l) = \frac{1}{2}\frac{\log \, \langle R^2(l+1, N) \rangle / \langle R^2(l, N) \rangle}{\log \, (l+1)/l}$ reported in the l.h.s insets of Fig.~\ref{fig:PathR2_ContactFreq}.

Then, we have calculated the mean closure probabilities, $\langle p_c(l, N) \rangle$ (Fig.~\ref{fig:PathR2_ContactFreq}, right-hand panels),
normalised to the corresponding ``mean-field'' expectation values $\langle R^2(l, N) \rangle^{-3/2} \sim l^{-3\nu_{\mathrm{path}}}$.
As in the case of single self-avoiding trees in good solvent~\cite{Rosa2016a}, $\langle p_c(l, N) \rangle$ for interacting trees markedly deviate from the mean-field prediction,
which defines a {\it novel} critical exponent $\theta_{\mathrm{path}}$,
$\langle p_c(l, N) \rangle \langle R^2(l, N) \rangle^{3/2} \sim l^{-\nu_{\mathrm{path}}\theta_{\mathrm{path}}}$.
Estimated values for $\theta_{\mathrm{path}}$'s are reported in Table~\ref{tab:ExpSummary}.

\subsection{Conformational statistics of trees}\label{sec:ConfStatTrees}

\begin{figure}
\begin{center}
\includegraphics[width=0.5\textwidth]{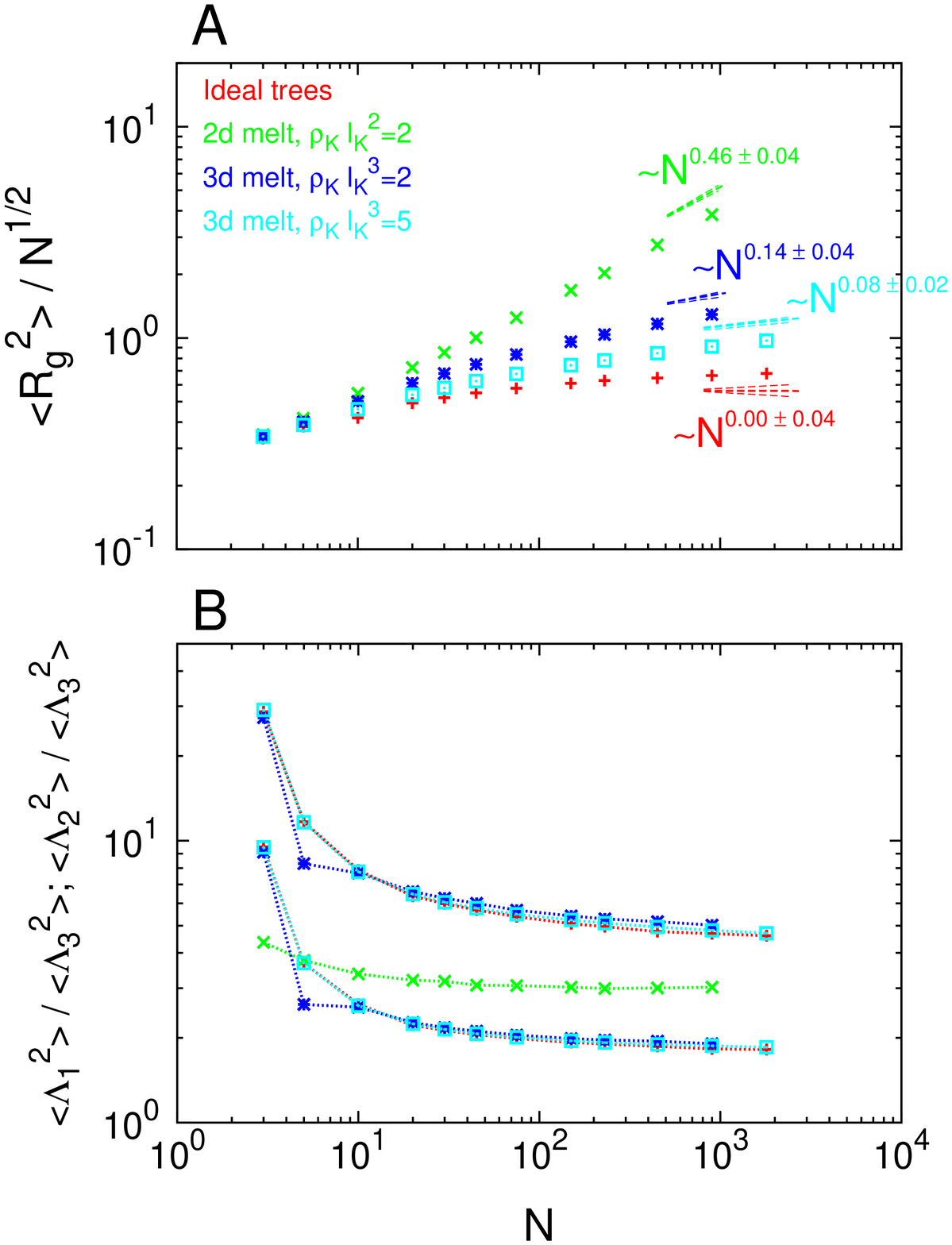}
\end{center}
\caption{
\label{fig:Rg2AspectRatios}
Conformational statistics of trees.
(A)
Mean square gyration radius, $\langle R_g^2 (N) \rangle$.
Straight lines correspond to the large-$N$ behaviour $\langle R_g^2(N) \rangle \sim N^{2 \nu}$ with critical exponents $\nu$
given by the best estimates summarised in Table~\ref{tab:ExpSummary}.
(B)
Tree average aspect ratios, $\langle \Lambda_1^2 \rangle / \langle \Lambda_3^2 \rangle$ and $\langle \Lambda_2^2 \rangle / \langle \Lambda_3^2 \rangle$.
}
\end{figure}

Finally,
we have measured the mean-square gyration radius, $\langle R_g^2(N) \rangle$ (tabulated values in Tables~SV-VI~\cite{SupplMatNote}), 
and the average shape of trees as a function of tree weight, $N$
(Fig.~\ref{fig:Rg2AspectRatios}, panels A and B respectively).
Estimated values of critical exponents $\nu$ (straight lines in Fig.~\ref{fig:Rg2AspectRatios}A) are summarized in Table~\ref{tab:ExpSummary},
while details about their derivation are given in Table~SVII~\cite{SupplMatNote}. 

\section{Discussion}\label{sec:Discussion}

\begin{table*}
\begin{tabular}{|c|c|c|c|c|c|c|c|cc|c|}
\cline{1-8} \cline{11-11}
& & \multicolumn{2}{c|}{\scriptsize Ideal trees with} & \multicolumn{2}{c|}{\scriptsize $2d$ melt of trees with} & \multicolumn{2}{c|}{\scriptsize $3d$ melt of trees with} & & & {\scriptsize $3d$ melt of trees with}\\
& & \multicolumn{2}{c|}{\scriptsize annealed connectivity} & \multicolumn{2}{c|}{\scriptsize annealed connectivity} & \multicolumn{2}{c|}{\scriptsize annealed connectivity} & & & {\scriptsize annealed connectivity}\\
\cline{1-8} \cline{11-11}
& {\scriptsize Relation to} & {\scriptsize Theory} & {\scriptsize Simul.} & {\scriptsize Theory} & {\scriptsize Simul., $\rho_K l_K^2 = 2$} & {\scriptsize Theory} & {\scriptsize Simul., $\rho_K l_K^3 = 2$} & & & {\scriptsize Simul., $\rho_K l_K^3 = 5$} \\
& {\scriptsize other exponents} & & & & & & & & & \\
\cline{1-8} \cline{11-11}
& & & & & & & & & & \\
{\scriptsize $\rho$} & -- & {\scriptsize $\frac{1}{2} = 0.5$} & {\scriptsize $0.49 \pm 0.04$} & {\scriptsize $\frac{2}{3} \approx 0.667$} & {\scriptsize $0.613 \pm 0.007$} & {\scriptsize $\frac{5}{9} \approx 0.556$} & {\scriptsize $0.52 \pm 0.05$} & & & {\scriptsize $0.53 \pm 0.03$} \\
& & & & & & & & & & \\
{\scriptsize $\epsilon$} & {\scriptsize $=\rho$} & {\scriptsize $\frac{1}{2} = 0.5$} & {\scriptsize $0.536 \pm 0.007$} & {\scriptsize $\frac{2}{3} \approx 0.667$} & {\scriptsize $0.63 \pm 0.01$} & {\scriptsize $\frac{5}{9} \approx 0.556$} & {\scriptsize $0.57 \pm 0.02$} & & & {\scriptsize $0.554 \pm 0.006$} \\
& & & & & & & & & & \\
{\scriptsize $\nu_{\mathrm{path}}$} & -- & {\scriptsize $\frac{1}{2} = 0.5$} & {\scriptsize $0.509 \pm 0.009$} & {\scriptsize $\frac{3}{4} = 0.75$} & {\scriptsize $0.780 \pm 0.005$} & {\scriptsize $\frac{3}{5} = 0.6$} & {\scriptsize $0.593 \pm 0.007$} & & & {\scriptsize $0.565 \pm 0.007$} \\
& & & & & & & & & & \\
{\scriptsize $\nu$} & {\scriptsize $= \rho \, \nu_{\mathrm{path}}$} & {\scriptsize $\frac{1}{4} = 0.25$} & {\scriptsize $0.25 \pm 0.02$} & {\scriptsize $\frac{1}{2} = 0.5$} & {\scriptsize $0.48 \pm 0.02$} & {\scriptsize $\frac{1}{3} \approx 0.333$} & {\scriptsize $0.32 \pm 0.02$} & & & {\scriptsize $0.29 \pm 0.01$} \\
& & & & & & & & & & \\
{\scriptsize $\theta_{\mathrm{path}}$} & -- & -- & {\scriptsize $-0.04 \pm 0.04$} & -- & {\scriptsize $0.65 \pm 0.06$} & -- & {\scriptsize $0.12 \pm 0.05$} & & & {\scriptsize $0.14 \pm 0.02$} \\
& & & & & & & & & & \\
\cline{1-8} \cline{11-11}
\end{tabular}
\caption{
\label{tab:ExpSummary}
{\it Asymptotic} (left) and {\it crossover} (right) critical exponents for $2d$ and $3d$ lattice trees.
(a)
$\rho$, $\epsilon$, $\nu_{\mathrm{path}}$ and $\nu$:
Comparison between predictions of Flory theory and numerical results.
As explained in Sec.~\ref{sec:ExtractExps}, numerical results for $\rho$, $\epsilon$ and $\nu$ (for this last exponent, with the only exception of $2d$ melts) were derived by combining values and statistical errors of the parameters obtained from best fits (Tables~SIV and~SVII~\cite{SupplMatNote}) of Eqs. (\ref{eq:FitRg22}) and~(\ref{eq:FitRg21lin2}) to corresponding observables.
For $\nu_{\mathrm{path}}$ and $\nu$ in $2d$ melts, Eq.~(\ref{eq:FitRg21lin2}) fails (Table~SVII~\cite{SupplMatNote}) and only Eq.~(\ref{eq:FitRg22}) was employed.
(b)
$\theta_{\mathrm{path}}$:
Numerical results are obtained by combining the estimated values of $\nu_{\mathrm{path}}$
with the values of ``$\nu_{\mathrm{path}} \theta_{\mathrm{path}}$''
averaged over the ranges of $l$'s where this quantity shows a quasi-plateau for $N=900$ or $N=1800$ 
(see grey and black lines in the insets of the r.h.s panels of Fig.~\ref{fig:PathR2_ContactFreq}):
$\nu_{\mathrm{path}}\theta_{\mathrm{path}} = -0.02 \pm 0.02$ (ideal trees);
$\nu_{\mathrm{path}}\theta_{\mathrm{path}} = 0.51 \pm 0.05$ ($2d$ melt of trees, $\rho_K l_K^2 = 2$);
$\nu_{\mathrm{path}}\theta_{\mathrm{path}} = 0.07 \pm 0.03$ ($3d$ melt of trees, $\rho_K l_K^3 = 2$);
$\nu_{\mathrm{path}}\theta_{\mathrm{path}} = 0.08 \pm 0.01$ ($3d$ melt of trees, $\rho_K l_K^3 = 5$).
Average values and corresponding error bars have been rounded to the first significant decimal digit.
}
\end{table*}

We have analyzed the behavior of interacting trees with annealed connectivity in $2d$ and $3d$ melts in terms of a small set of exponents defined in the Introduction and in Sec.~\ref{sec:observables}.
With weights of up to $N=1800$ segments, our tree sizes are comparable to those of linear chains in similar studies~\cite{MeyerWittmerJCP2010,WittmerReview2011,MeyerWittmerMacromolecules2012}. 
However, for many purposes the average contour distance, $\langle L\rangle \le {\cal O}(100)\ll N$, between monomers provides a more suitable comparison.
The reader should thus bear in mind, that extracted exponents are either (i) effective (crossover) exponents valid for the particular systems and system sizes we have studied or (ii) {\em estimates} of true,
asymptotic exponents, which suffer from uncertainties related to the extrapolation to the asymptotic limit. 
These effects are particularly pronounced for the high density $3d$ melts we studied in Ref.~\cite{RosaEveraersPRL2014}. For example, we initially reported~\cite{RosaEveraersPRL2014} the results of a simple power law fit  including  all data for tree sizes $N \geq 75$ suggesting $\nu=0.32 \pm 0.01$. In contrast, the present more refined analysis yields $\nu=0.29 \pm 0.01$. If this result is acceptable as an effective exponent for tree sizes $N={\mathcal O}(1000)$, it mostly illustrates that one cannot extract the asymptotic behaviour assuming a {\em single}, small correction to scaling, if the available tree sizes barely reach the crossover (Sec.~\ref{sec:BlobSize2}).  
The discussion below focuses on results for lower density $2d$ and $3d$ melts, where the studied tree sizes are significantly larger than the estimated blob size.  

Our results are summarised in Table~\ref{tab:ExpSummary}.
Most of them are in quantitative or at least good qualitative agreement with the predictions from Flory theory~\cite{IsaacsonLubensky,DaoudJoanny1981,GutinGrosberg93,GrosbergSoftMatter2014}.
First,
this implies that the reported exponents $\nu_{\mathrm{path}}$ for melts of trees fit to the predicted~\cite{GrosbergSoftMatter2014} value $\frac{3}{d+2}$
which, intriguingly, corresponds to the Flory~\cite{Flory1969} exponent $\nu$ for self-avoiding {\it linear} chains in good solvent conditions.
Second, 
the results for the exponent $\nu$ are in agreement with the prediction $\nu=1/d$ and confirm that trees in melts behave as ``territorial'' polymers. 
This result is relatively simple to understand as $\nu=1/d$ represents the minimal amount of swelling compatible with steric packing requirements.
Thus not too much stock should be put in the fact that the Flory value for $\nu$ turns out to be exact in the present case.
More interestingly, the theory makes two testable, non-trivial predictions for the contributions of connectivity changes and path stretching to the overall swelling~\cite{GutinGrosberg93} (Eqs.~(\ref{eq:rho_of_nu}) and (\ref{eq:nupath_of_nu})):
firstly, both effects should only depend on the magnitude of the overall swelling,
but not its physical origin and secondly path stretching is expected to be dominant as~\cite{WeakStretchingNote}
$\nu_{\mathrm{path}}-\nu_{\mathrm{path}}^{\mathrm{ideal}} \approx 2(\rho-\rho^{\mathrm{ideal}})=4(\nu-\nu^{\mathrm{ideal}} )/3$ for weakly swollen trees. 
The first prediction is well borne out by the comparison of $3d$ self avoiding trees and $2d$ melt trees.
In both cases, the trees are expected to swell to $\nu=1/2$. Interestingly, the observed values for $\rho$ and $\nu_{\mathrm{path}}$ turn out to be almost identical and close to the predicted values. 
The second prediction is confirmed by noticing that trees in $2d$ and $3d$ melts 
swell almost exclusively at the path level (Fig.~\ref{fig:R2MaxLmax}), while the observed modifications of the connectivities
(Figs.~\ref{fig:PathLengthStat_L} and~\ref{fig:nBranch}), are very weak.
In absolute terms for our largest trees with $N=900$
(see Tables~SII-III~\cite{SupplMatNote} and SV-VI~\cite{SupplMatNote}):
$\langle R_g^2\rangle_{2d\ \mathrm{melt}} / \langle R_g^2\rangle_{\mathrm{ideal}} = 5.78\pm0.03$, 
$\langle R^2(L_{\mathrm{max}})\rangle_{2d\ \mathrm{melt}} / \langle R^2(L_{\mathrm{max}})\rangle_{\mathrm{ideal}} = 6.32\pm0.10$ and
$\langle L\rangle_{2d\ \mathrm{melt}} / \langle L\rangle_{\mathrm{ideal}} = 1.188\pm0.002$ while
$\langle R_g^2\rangle_{3d\ \mathrm{melt}} / \langle R_g^2\rangle_{\mathrm{ideal}} = 1.95\pm0.02$, 
$\langle R^2(L_{\mathrm{max}})\rangle_{3d\ \mathrm{melt}} / \langle R^2(L_{\mathrm{max}})\rangle_{\mathrm{ideal}} = 1.96\pm0.03$ and
$\langle L\rangle_{3d\ \mathrm{melt}} / \langle L\rangle_{\mathrm{ideal}} = 1.054\pm0.003$.

For other quantities, Flory theory is even qualitatively wrong. 
A particularly interesting case are the average contact probability $\langle p_c(l) \rangle$ between nodes at path distance $l$.
As shown in Fig.~\ref{fig:PathR2_ContactFreq} and Table~\ref{tab:ExpSummary},
$\langle p_c(l) \rangle$'s for interacting trees deviate consistently from the na\"ive mean-field estimate of $l^{-3\nu_{\mathrm{path}}}$.
This is yet another illustration of the subtle cancellation of errors in Flory arguments, which are built on the mean-field estimates of contact probabilities~\cite{DeGennesBook}.

\section{Summary and Conclusion}\label{sec:concls}


Motivated by the close analogy between non-concatenated ring polymers and annealed lattice trees~\cite{KhokhlovNechaev85,RubinsteinPRL1986,RubinsteinPRL1994,RosaEveraersPRL2014,GrosbergSoftMatter2014}, we have studied the statistical properties of tree melts in $d=2$ and $d=3$ dimensions. 
We have used the same methodology as in our recent work on self-avoiding trees~\cite{Rosa2016a}, {\it i.e.} variants of the amoeba~\cite{SeitzKlein1981} and burning~\cite{StanleyJPhysA1984,StaufferAharonyBook} algorithms for the Monte Carlo simulation and the connectivity analysis (Sec.~\ref{sec:modmethods}).
Table~\ref{tab:ExpSummary} summarises our estimates of the asymptotic values of the exponents describing the scaling behavior of 
the average branch weight, $\langle N_{br} (N) \rangle \sim N^{\epsilon}$,
the average path length, $\langle L(N) \rangle \sim N^{\rho}$,
the mean-square path extension, $\langle R^2(l) \rangle \sim l^{2\nu_{\mathrm{path}}}$,
and the tree and branch gyration radii, $\langle R_g^2(N) \rangle \sim N^{2\nu}$ (Sec.~\ref{sec:results}).
Our results are in excellent agreement with an asymptotic scaling of the average tree size of  $R \sim N^{1/d}$, suggesting that the trees behave as compact, {\it territorial}~\cite{VettorelPhysToday2009} fractals (Fig.~\ref{fig:Rg2AspectRatios}). 
Moreover, we find that the trees swell by the combination of modified branching and path stretching.
However, the former effect is subdominant and difficult to detect in $d=3$ dimensions.

Our results for dense systems contribute to the evidence 
suggesting that Flory theory~\cite{IsaacsonLubensky,DaoudJoanny1981,GutinGrosberg93,GrosbergSoftMatter2014}
provides a useful framework for discussing the behavior of interacting trees.
That Flory theory should work for trees is not a foregone conclusion.
In the case of linear chains, the approach is notorious (and appreciated) for the nearly perfect cancellation of large errors in the estimation of {\em both} terms in Eq.~(\ref{eq:fFlory}) \cite{DeGennesBook,DesCloizeauxBook}.
This delicate balance might well have been destroyed for trees, where the Flory energy needs to be simultaneously minimized with respect to $L$ and $R$.
In two forthcoming publications, we will generalise Flory theory to trees of finite size and extensibility~\cite{Everaers2016b} and we will analyse the distribution functions for the quantities,
whose {\em mean} behaviour we have explored above and in Ref.~\cite{Rosa2016a} in an attempt to go beyond Flory theory~\cite{Rosa2016c}.

{\it Acknowledgements} --
AR acknowledges grant PRIN 2010HXAW77 (Ministry of Education, Italy).
RE is grateful for the hospitality of the Kavli Institute for Theoretical Physics (Santa Barbara, USA)
and support through the National Science Foundation under Grant No. NSF PHY11-25915 during his visit in 2011, which motivated the present work. In particular, we have benefitted from long and stimulating exchanges with M. Rubinstein and A. Yu. Grosberg on the theoretical background.
This work was only possible thanks to generous grants of computer time by PSMN (ENS-Lyon) and P2CHPD (UCB Lyon 1), in part through the equip@meso facilities of the FLMSN.


\begin{thebibliography}{10}

\bibitem{IsaacsonLubensky}
J.~Isaacson and T.~C. Lubensky.
\newblock Flory exponents for generalized polymer problems.
\newblock {\em J. Physique Lett.}, 41:L469--L471, 1980.

\bibitem{SeitzKlein1981}
W.~A. Seitz and D.~J. Klein.
\newblock Excluded volume effects for branched polymers.
\newblock {\em J. Chem. Phys.}, 75:5190--5193, 1981.

\bibitem{DuarteRuskin1981}
J.~A. M.~S. Duarte and H.~J. Ruskin.
\newblock The branching of real lattice trees as dilute polymers.
\newblock {\em J. Physique}, 42:1585--1590, 1981.

\bibitem{ParisiSourlasPRL1981}
G.~Parisi and N.~Sourlas.
\newblock Critical behavior of branched polymers and the {L}ee-{Y}ang edge
  singularity.
\newblock {\em Phys. Rev. Lett.}, 46:871--874, 1981.

\bibitem{FisherPRL1978}
M.~E. Fisher.
\newblock Yang-{L}ee edge singularity and $\phi^3$ field-theory.
\newblock {\em Phys. Rev. Lett.}, 40:1610--1613, 1978.

\bibitem{KurtzeFisherPRB1979}
D.~A. Kurtze and M.~E. Fisher.
\newblock Yang-{L}ee edge singularities at high-temperatures.
\newblock {\em Phys. Rev. B}, 20:2785--2796, 1979.

\bibitem{BovierFroelichGlaus1984}
A.~Bovier, J.~Fr{\"o}hlich, and U.~Glaus.
\newblock {\em ``Branched Polymers and Dimensional Reduction'' in ``Critical
  Phenomena, Random Systems, Gauge Theories''}.
\newblock North-Holland, Amsterdam, K. Osterwalder and R. Stora (eds.), 1984.

\bibitem{Read2013}
P.~Bacova, L.~G.~D. Hawke, D.~J. Read, and A.~J. Moreno.
\newblock Dynamics of branched polymers: A combined study by molecular dynamics
  simulations and tube theory.
\newblock {\em Macromolecules}, 46:4633--4650, 2013.

\bibitem{RubinsteinColby}
M.~Rubinstein and R.~H. Colby.
\newblock {\em Polymer Physics}.
\newblock Oxford University Press, New York, 2003.

\bibitem{Burchard1999}
W.~Burchard.
\newblock Solution properties of branched macromolecules.
\newblock {\em Adv. Polym. Sci.}, 143:113, 1999.

\bibitem{KhokhlovNechaev85}
A.~R. Khokhlov and S.~K. Nechaev.
\newblock Polymer chain in an array of obstacles.
\newblock {\em Phys. Lett.}, 112A:156--160, 1985.

\bibitem{RubinsteinPRL1986}
M.~Rubinstein.
\newblock Dynamics of ring polymers in the presence of fixed obstacles.
\newblock {\em Phys. Rev. Lett.}, 57:3023--3026, 1986.

\bibitem{RubinsteinPRL1994}
S.~P. Obukhov, M.~Rubinstein, and T.~Duke.
\newblock Dynamics of a ring polymer in a gel.
\newblock {\em Phys. Rev. Lett.}, 73:1263--1266, 1994.

\bibitem{kapnistos2008}
M.~Kapnistos et~al.
\newblock Unexpected power-law stress relaxation of entangled ring polymers.
\newblock {\em Nature Materials}, 7:997, 2008.

\bibitem{GrosbergSoftMatter2014}
A.~Yu. Grosberg.
\newblock Annealed lattice animal model and flory theory for the melt of
  non-concatenated rings: Towards the physics of crumpling.
\newblock {\em Soft Matter}, 10:560--565, 2014.

\bibitem{RosaEveraersPRL2014}
A.~Rosa and R.~Everaers.
\newblock Ring polymers in the melt state: The physics of crumpling.
\newblock {\em Phys. Rev. Lett.}, 112:118302, 2014.

\bibitem{Rosa2016a}
A.~Rosa and R.~Everaers.
\newblock Computer simulations of randomly branching polymers: Annealed vs.
  quenched branching structures.
\newblock {\em J. Phys. A: Math. Theor.}, 49:345001, 2016.

\bibitem{grosbergEPL1993}
A.~Grosberg, Y.~Rabin, S.~Havlin, and A.~Neer.
\newblock Crumpled globule model of the three-dimensional structure of {DNA}.
\newblock {\em Europhys. Lett.}, 23:373--378, 1993.

\bibitem{RosaPLOS2008}
A.~Rosa and R.~Everaers.
\newblock Structure and dynamics of interphase chromosomes.
\newblock {\em Plos Comput. Biol.}, 4:e1000153, 2008.

\bibitem{Vettorel2009}
T.~Vettorel, A.~Y. Grosberg, and K.~Kremer.
\newblock Statistics of polymer rings in the melt: a numerical simulation
  study.
\newblock {\em Phys. Biol.}, 6:025013, 2009.

\bibitem{MirnyRev2011}
Leonid~A. Mirny.
\newblock The fractal globule as a model of chromatin architecture in the cell.
\newblock {\em Chromosome Res.}, 19:37--51, 2011.

\bibitem{LangMacromol2013}
M.~Lang.
\newblock Ring conformations in bidisperse blends of ring polymers.
\newblock {\em Macromolecules}, 46:1158--1166, 2013.

\bibitem{SmrekGrosbergACSMacroLett2016}
J.~Smrek and A.~Yu. Grosberg.
\newblock Minimal surfaces on unconcatenated polymer rings in melt.
\newblock {\em ACS Macro Letters}, 5:750--754, 2016.

\bibitem{CatesDeutsch}
M.~E. Cates and J.~M. Deutsch.
\newblock Conjectures on the statistics of ring polymers.
\newblock {\em J. Phys. (Paris)}, 47:2121--2128, 1986.

\bibitem{RubinsteinMacromolecules2016}
T.~Ge, S.~Panyukov, and M.~Rubinstein.
\newblock Self-similar conformations and dynamics in entangled melts and
  solutions of nonconcatenated ring polymers.
\newblock {\em Macromolecules}, 49:708--722, 2016.

\bibitem{DeGennesBook}
P.-G. {De Gennes}.
\newblock {\em Scaling Concepts in Polymer Physics}.
\newblock Cornell University Press, Ithaca, 1979.

\bibitem{DoiEdwards}
M.~Doi and S.~F. Edwards.
\newblock {\em The Theory of Polymer Dynamics}.
\newblock Oxford University Press, New York, 1986.

\bibitem{KhokhlovGrosberg}
A.~Yu. Grosberg and A.~R. Khokhlov.
\newblock {\em Statistical Physics of Macromolecules}.
\newblock AIP Press, New York, 1994.

\bibitem{ZimmStockmayer49}
B.~H. Zimm and W.~H. Stockmayer.
\newblock The dimensions of chain molecules containing branches and rings.
\newblock {\em J. Chem. Phys.}, 17:1301--1314, 1949.

\bibitem{DaoudJoanny1981}
M.~Daoud and J.~F. Joanny.
\newblock Conformation of branched polymers.
\newblock {\em J. Physique}, 42:1359--1371, 1981.

\bibitem{GutinGrosberg93}
A.~M. Gutin, A.~Yu. Grosberg, and E.~I. Shakhnovich.
\newblock Polymers with annealed and quenched branchings belong to different
  universality classes.
\newblock {\em Macromolecules}, 26:1293--1295, 1993.

\bibitem{DesCloizeauxBook}
J.~des Cloizeaux and G.~Jannink.
\newblock {\em Polymers in Solution}.
\newblock Oxford University Press, Oxford, 1989.

\bibitem{Everaers2016a}
R.~Everaers, A.~Yu. Grosberg, M.~Rubinstein, and A.~Rosa.
\newblock Flory theory of randomly branching polymers {I}: Asymptotic
  behaviour.
\newblock {\em In preparation}, 2016.

\bibitem{Everaers2016b}
R.~Everaers.
\newblock Flory theory of randomly branching polymers {II}: Interacting trees
  of finite size and finite extensibility.
\newblock {\em In preparation}, 2016.

\bibitem{Rosa2016c}
A.~Rosa and R.~Everaers.
\newblock Beyond {F}lory theory: Distribution functions for interacting lattice
  trees.
\newblock {\em In preparation}, 2016.

\bibitem{MadrasJPhysA1992}
E.~J. {Janse van Rensburg} and N.~Madras.
\newblock A nonlocal monte carlo algorithm for lattice trees.
\newblock {\em J. Phys. A: Math. Gen.}, 25:303--333, 1992.

\bibitem{FloryChemBook}
P.~J. Flory.
\newblock {\em Principles of Polymer Chemistry}.
\newblock Cornell University Press, Ithaca (NY), 1953.

\bibitem{GrosbergNechaev2015}
A.~Yu. Grosberg and S.~K. Nechaev.
\newblock From statistics of regular tree-like graphs to distribution function
  and gyration radius of branched polymers.
\newblock {\em J. Phys. A-Math. Theor.}, 48:345003, 2015.

\bibitem{BlobSizeNote}
The result is consistent with the ``traditional''
  picture~\cite{RubinsteinColby} that polymers in semi-dilute solutions can be
  represented as chains of blobs, where each blob contains $g$ monomers and has
  a linear size $R(g) \sim l_K g^\nu$. Since blobs are densily
  packed~\cite{RubinsteinColby}, the monomer density inside each blob equals
  the overall density or, neglecting prefactors, $\rho_K \sim g/R(g)^3
  \rightarrow g \sim (\rho_K l_K^3)^{1/(1-3\nu)}$. For standard semi-dilute
  solutions of {\it linear} chains, $\nu \approx 0.6$ and $g \sim (\rho_K
  l_K^3)^{-1.25}$ {\it decreases} with density~\cite{RubinsteinColby}. For our
  melts of trees, $\nu=1/4$ (see Fig.~\ref{fig:Rg2AspectRatios}A) and $g \sim
  (\rho_K l_K^3)^4$ (equivalent within numerical prefactors to
  Eq.~(\ref{eq:blob3d})) which {\it increases} with density. This picture ought
  to break when density fluctuations (see discussion at the end of
  Sec.~\ref{sec:ModelUnits}) become comparable to $\rho_K$, or $\rho_K l_K^3
  \approx \sqrt{0.13} \approx 0.4$.

\bibitem{StanleyJPhysA1984}
H.~J. Heermann, D.~C. Hong, and H.~E. Stanley.
\newblock Backbone and elastic backbone of percolation clusters obtained by the
  new method of `burning'.
\newblock {\em J. Phys. A: Math. Gen.}, 17:L261--L266, 1984.

\bibitem{StaufferAharonyBook}
D.~Stauffer and A.~Aharony.
\newblock {\em Introduction to percolation theory}.
\newblock Taylor \& Francis Inc., 1994.

\bibitem{SupplMatNote}
See supplemental material at [URL] for complementary figures and tables cited
  in the article.

\bibitem{NumericalRecipes}
W.~H. Press, S.~A. Teukolsky, W.~T. Vetterling, and B.~F. Flannery.
\newblock {\em Numerical Recipes in Fortran}.
\newblock Cambridge University Press, Cambridge, 2 edition, 1992.

\bibitem{MeyerWittmerJCP2010}
H.~Meyer, J.~P. Wittmer, T.~Kreer, A.~Johner, and J.~Baschnagel.
\newblock {Static properties of polymer melts in two dimensions}.
\newblock {\em {J. Chem. Phys.}}, {132}:{184904}, {2010}.

\bibitem{WittmerReview2011}
J.~P. Wittmer et~al.
\newblock Scale-free static and dynamical correlations in melts of monodisperse
  and flory-distributed homopolymers.
\newblock {\em J. Stat. Phys.}, 145:1017--1126, 2011.

\bibitem{MeyerWittmerMacromolecules2012}
N.~Schulmann, H.~Meyer, J.~P. Wittmer, A.~Johner, and J.~Baschnagel.
\newblock {Interchain Monomer Contact Probability in Two-Dimensional Polymer
  Solutions}.
\newblock {\em {Macromolecules}}, {45}:{1646--1651}, {2012}.

\bibitem{Flory1969}
P.~J. Flory.
\newblock {\em Statistical Mechanics of Chain Molecules}.
\newblock Interscience, New York, 1969.

\bibitem{WeakStretchingNote}
The result is trivially obtained by expanding Eqs.~(\ref{eq:rho_of_nu})
  and~(\ref{eq:nupath_of_nu}) around the ideal values
  $\nu_{\mathrm{path}}^{\mathrm{ideal}} = \rho_{\mathrm{path}}^{\mathrm{ideal}}
  = 1/2$ and $\nu^{\mathrm{ideal}}=1/4$.

\bibitem{VettorelPhysToday2009}
T.~Vettorel, A.~Y. Grosberg, and K.~Kremer.
\newblock Territorial polymers.
\newblock {\em Phys. Today}, 62:72, 2009.

\end{thebibliography}

\clearpage

\setcounter{section}{0}
\setcounter{figure}{0}
\setcounter{table}{0}
\setcounter{equation}{0}

\renewcommand{\figurename}{Fig. S}
\renewcommand{\tablename}{Table S}

{\large \bf Supplemental Material}

\tableofcontents

\clearpage

\section{Supplementary Figures}

\begin{figure*}
\includegraphics[width=0.5\textwidth]{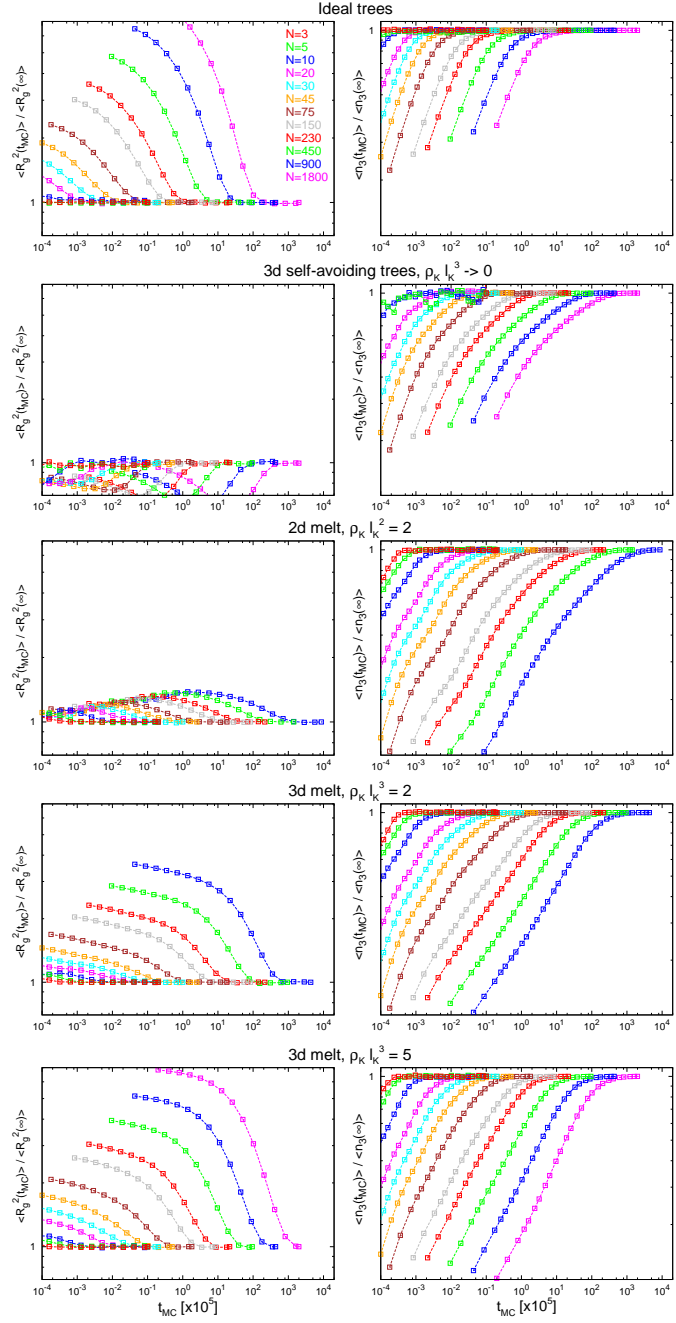}
\caption{
\label{fig:MC_Equilibration_RgN3}
Equilibration of $2d$ and $3d$ lattice trees of mass size $N$ via the ``amoeba'' algorithm (Sec.~IIIA, main paper) for different ensembles, as a function of Monte Carlo time-steps ($t_{MC}$).
Left and right panels are for, respectively, the (ensemble-average) square gyration radius ($\langle R_g^2(t_{MC}) \rangle$) and number of branching nodes ($\langle n_3(t_{MC}) \rangle$) normalized to the corresponding asymptotic values.
For comparison, data for $3d$ self-avoiding trees ({\it i.e.}, zero density) from [Rosa \& Everaers, {\it J. Phys. A.: Math. Theor.} {\bf 49}, 345001 (2016)] are also reported.
In all cases, the initial configuration for each tree corresponds to a linearly connected random walk.
}
\end{figure*}

\begin{figure*}
\includegraphics[width=0.5\textwidth]{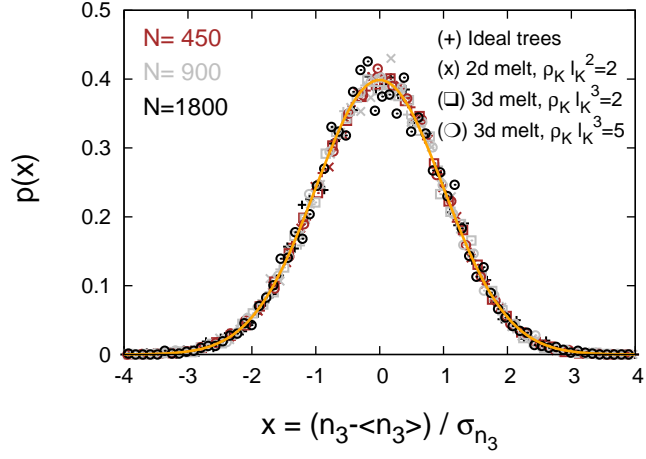}
\caption{
\label{fig:3fDistrs}
Branching statistics.
Distribution functions for the number of branching points $n_3$ (symbols) follow the Gaussian distribution (orange solid line).
Corresponding variances $\sigma_{n_3}^2$ increase linearly with $N$ as:
$\sigma_{n_3}^2 / N = 0.0410 \pm 0.0002$ (ideal trees);
$\sigma_{n_3}^2 / N =  0.0447 \pm 0.0002$ ($2d$ melt of trees, $\rho_K l_K^2 = 2$);
$\sigma_{n_3}^2 / N = 0.0428 \pm 0.0003$ ($3d$ melt of trees, $\rho_K l_K^3 = 2$);
$\sigma_{n_3}^2 / N = 0.0424 \pm 0.0002$ ($3d$ melt of trees, $\rho_K l_K^3 = 5$).
}
\end{figure*}

\begin{figure*}
\includegraphics[width=\textwidth]{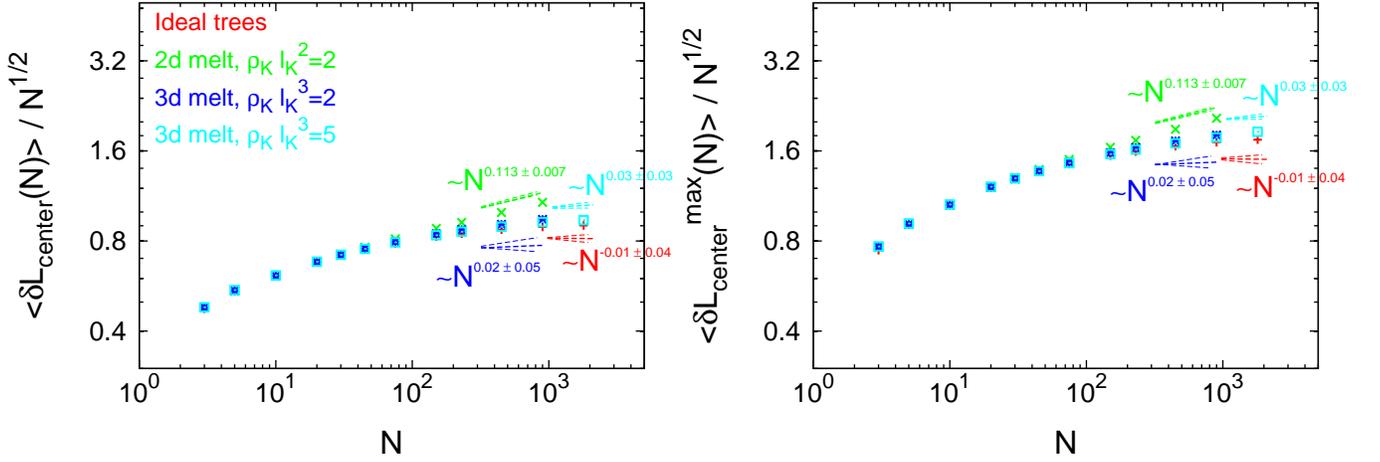}
\caption{
\label{fig:PathLengthStat_dLcenterdLcenterMax}
Path length statistics.
(Left)
Average contour distance of nodes from the central node, $\langle \delta L_{\mathrm{center}}(N) \rangle$.
(Right)
Average longest contour distance of nodes from the central node, $\langle \delta L_{\mathrm{center}}^{\mathrm{max}}(N) \rangle$.
Straight lines correspond to the large-$N$ behaviour
$\langle \delta L_{\mathrm{center}}(N) \rangle \sim \langle \delta L_{\mathrm{center}}^{\mathrm{max}}(N) \rangle \sim N^{\rho}$
with critical exponents $\rho$ given by the best estimates summarised in Table~I 
(main paper) and Table~S\ref{tab:FitsCritExpsTreeStructure}. 
}
\end{figure*}

\clearpage

\section{Supplementary Tables}

\begin{table*}
\begin{tabular}{|c|c|c|c|c|c|c|c|}
\hline
\multicolumn{2}{|c}{} & \multicolumn{3}{|c}{$2d$ ideal trees} & \multicolumn{3}{|c|}{$3d$ ideal trees} \\
\hline
{$N$} & {$M$} & {$R$} & {$\tau_{MC}$} & {$\tau_{MC} / \tau_{corr}$} & {$R$} & {$\tau_{MC}$} & {$\tau_{MC} / \tau_{corr}$}\\
\hline
{3} & {1} & {$12800$} & {$5 \cdot 10^3$} & {$\approx 833$} & {$16000$} & {$1 \cdot 10^4$} & {$\approx 1500$}\\
{5} & {1} & {$11200$} & {$5 \cdot 10^3$} & {$\approx 208$} & {$ 6400$} & {$1 \cdot 10^4$} & {$\approx  400$}\\
{10} & {1} & {$12000$} & {$5 \cdot 10^3$} & {$\approx 52$} & {$ 3200$} & {$1 \cdot 10^4$} & {$\approx  100$}\\
{20} & {1} & {$12000$} & {$5 \cdot 10^3$} & {$\approx 13$} & {$25000$} & {$1 \cdot 10^4$} & {$\approx   25$}\\
{30} & {1} & {$12800$} & {$1 \cdot 10^4$} & {$\approx 10$} & {$25600$} & {$2 \cdot 10^4$} & {$\approx   20$}\\
{45} & {1} & {$32000$} & {$2.5 \cdot 10^4$} & {$\approx 10$} & {$25600$} & {$5 \cdot 10^4$} & {$\approx   20$}\\
{75} & {1} & {$38400$} & {$1 \cdot 10^5$} & {$\approx 10$} & {$25600$} & {$2 \cdot 10^5$} & {$\approx   20$}\\
{150} & {1} & {$30000$} & {$5 \cdot 10^5$} & {$\approx 7$} & {$25600$} & {$1 \cdot 10^6$} & {$\approx   13$}\\
{230} & {1} & {$32000$} & {$1 \cdot 10^6$} & {$\approx 7$} & {$25600$} & {$2 \cdot 10^6$} & {$\approx   13$}\\
{450} & {1} & {$32000$} & {$5 \cdot 10^6$} & {$\approx 7$} & {$25600$} & {$1 \cdot 10^7$} & {$\approx   13$}\\
{900} & {1} & {$16000$} & {$2 \cdot 10^7$} & {$\approx 7$} & {$12800$} & {$4 \cdot 10^7$} & {$\approx   13$}\\
{1800} & {1} & {$8000$} & {$1 \cdot 10^8$} & {$\approx 7$} & {$ 6400$} & {$2 \cdot 10^8$} & {$\approx   13$}\\
\hline
\end{tabular}\\
\begin{tabular}{|c|c|c|c|c|c|c|c|c|c|c|c|c|}
\multicolumn{13}{c}{ }\\
\hline
& \multicolumn{4}{|c}{$2d$ melt of trees, $\rho_K l_K^2 = 2$} & \multicolumn{4}{|c}{$3d$ melt of trees, $\rho_K l_K^3 = 2$} & \multicolumn{4}{|c|}{$3d$ melt of trees, $\rho_K l_K^3 = 5$}\\
\hline
{$N$} & {$M$} & {$R$} & {$\tau_{MC}$} & {$\tau_{MC} / \tau_{corr}$} & {$M$} & {$R$} & {$\tau_{MC}$} & {$\tau_{MC} / \tau_{corr}$} & {$M$} & {$R$} & {$\tau_{MC}$} & {$\tau_{MC} / \tau_{corr}$}\\
\hline
{3} & {$128$} & {$100$} & {$2 \cdot 10^4$} & {$\approx 600$} & {$256$} & {$100$} & {$2 \cdot 10^4$} & {$\approx 800$} & {$160$} & {$100$} & {$1 \cdot 10^4$} & {$\approx 400$}\\
{5} & {$112$} & {$100$} & {$2 \cdot 10^4$} & {$\approx 150$} & {$256$} & {$100$} & {$2 \cdot 10^4$} & {$\approx 200$} & {$ 64$} & {$100$} & {$1 \cdot 10^4$} & {$\approx 150$}\\
{10} & {$120$} & {$100$} & {$2 \cdot 10^4$} & {$\approx 10$} & {$200$} & {$100$} & {$2 \cdot 10^4$} & {$\approx 20$} & {$ 32$} & {$100$} & {$1 \cdot 10^4$} & {$\approx  26$}\\
{20} & {$120$} & {$100$} & {$2 \cdot 10^4$} & {$\approx 3$} & {$288$} & {$100$} & {$2 \cdot 10^4$} & {$\approx 3$} & {$250$} & {$100$} & {$1 \cdot 10^4$} & {$\approx   3$}\\
{30} & {$128$} & {$100$} & {$1 \cdot 10^5$} & {$\approx 4$} & {$336$} & {$100$} & {$1 \cdot 10^5$} & {$\approx 6$} & {$256$} & {$100$} & {$2 \cdot 10^4$} & {$\approx   3$}\\
{45} & {$320$} & {$100$} & {$2.5 \cdot 10^5$} & {$\approx 3$} & {$256$} & {$100$} & {$2.5 \cdot 10^5$} & {$\approx 6$} & {$256$} & {$100$} & {$5 \cdot 10^4$} & {$\approx   3$}\\
{75} & {$384$} & {$100$} & {$1.8 \cdot 10^6$} & {$\approx 6$} & {$256$} & {$100$} & {$1.8 \cdot 10^6$} & {$\approx 10$} & {$256$} & {$100$} & {$2 \cdot 10^5$} & {$\approx   3$}\\
{150} & {$300$} & {$100$} & {$8.4 \cdot 10^6$} & {$\approx 4$} & {$288$} & {$100$} & {$8.4 \cdot 10^6$} & {$\approx 8$} & {$256$} & {$100$} & {$1 \cdot 10^6$} & {$\approx   2$}\\
{230} & {$288$} & {$100$} & {$2.2 \cdot 10^7$} & {$\approx 3$} & {$264$} & {$100$} & {$2.2 \cdot 10^7$} & {$\approx 8$} & {$256$} & {$100$} & {$2 \cdot 10^6$} & {$\approx   2$}\\
{450} & {$320$} & {$100$} & {$1.4 \cdot 10^8$} & {$\approx 3$} & {$256$} & {$100$} & {$1 \cdot 10^8$} & {$\approx 6$} & {$256$} & {$100$} & {$1 \cdot 10^7$} & {$\approx   2$}\\
{900} & {$88$} & {$100$} & {$8.7 \cdot 10^8$} & {$\approx 3$} & {$72$} & {$100$} & {$4 \cdot 10^8$} & {$\approx 6$} & {$128$} & {$100$} & {$4 \cdot 10^7$} & {$\approx   2$}\\
{1800} & -- & -- & -- & -- & -- & -- & -- & -- & {$ 64$} & {$100$} & {$2 \cdot 10^8$} & {$\approx   2$}\\
\hline
\end{tabular}
\caption{
\label{tab:MCruns}
Summary of MC results for trees from different ensembles.
$N$: number of Kuhn segments per tree.
$R$: total number of independent runs.
$M$: total number of trees per run.
$\tau_{MC}$: $\frac{\mbox{MC steps per single run}}{\mbox{single tree}}$.
$\tau_{MC} / \tau_{corr}$: total number of uncorrelated MC configurations per MC run.
$\tau_{corr}$ is the correlation time estimated {\it via} $g_3(\tau_{corr}) \approx \langle R_g^2\rangle$ (see also Fig.~2, main paper) 
where:
$g_3(\tau)$ is the mean-square displacement of the tree center of mass
as a function of MC steps 
and $\langle R_g^2\rangle$ is the tree mean-square gyration radius.
}
\end{table*}

\begin{table*}
\begin{tabular}{|c|c|c|c|c|c|}
\hline
{$N$} & {$\langle n_3\rangle$} & {$\langle \delta L_{\mathrm{center}} \rangle$} & {$\langle \delta L_{\mathrm{center}}^{\mathrm{max}} \rangle$} & {$\langle L \rangle$} & {$\langle N_{br} \rangle $}\\
\hline
\multicolumn{6}{c}{ }\\
\hline
\multicolumn{6}{|c|}{$2d$ ideal trees}\\
\hline
{3} & {$  0.718 \pm 0.004$} & {$ 0.821 \pm 0.001$} & {$ 1.282 \pm 0.004$} & {$ 1.160 \pm 0.001$} & {$ 0.094 \pm 0.001$}\\
{5} & {$  1.429 \pm 0.006$} & {$ 1.220 \pm 0.001$} & {$ 2.033 \pm 0.002$} & {$ 1.692 \pm 0.001$} & {$ 0.464 \pm 0.001$}\\
{10} & {$  3.334 \pm 0.006$} & {$ 1.932 \pm 0.002$} & {$ 3.335 \pm 0.004$} & {$ 2.763 \pm 0.001$} & {$ 1.093 \pm 0.002$}\\
{20} & {$  7.292 \pm 0.009$} & {$ 3.035 \pm 0.003$} & {$ 5.394 \pm 0.007$} & {$ 4.399 \pm 0.003$} & {$ 2.129 \pm 0.003$}\\
{30} & {$ 11.240 \pm 0.010$} & {$ 3.922 \pm 0.004$} & {$ 7.057 \pm 0.009$} & {$ 5.716 \pm 0.003$} & {$ 2.969 \pm 0.004$}\\
{45} & {$ 17.206 \pm 0.008$} & {$ 5.022 \pm 0.004$} & {$ 9.129 \pm 0.007$} & {$ 7.355 \pm 0.004$} & {$ 4.016 \pm 0.003$}\\
{75} & {$ 29.102 \pm 0.009$} & {$ 6.811 \pm 0.005$} & {$12.521 \pm 0.009$} & {$10.018 \pm 0.005$} & {$ 5.714 \pm 0.005$}\\
{150} & {$ 58.862 \pm 0.013$} & {$10.120 \pm 0.008$} & {$18.881 \pm 0.016$} & {$14.967 \pm 0.009$} & {$ 8.865 \pm 0.009$}\\
{230} & {$ 90.591 \pm 0.018$} & {$12.818 \pm 0.011$} & {$24.112 \pm 0.020$} & {$19.011 \pm 0.012$} & {$11.438 \pm 0.011$}\\
{450} & {$177.938 \pm 0.024$} & {$18.471 \pm 0.016$} & {$35.137 \pm 0.030$} & {$27.485 \pm 0.018$} & {$16.836 \pm 0.017$}\\
{900} & {$356.508 \pm 0.050$} & {$26.702 \pm 0.035$} & {$51.313 \pm 0.063$} & {$39.818 \pm  0.034$} & {$24.691 \pm 0.035$}\\
{1800} & {$713.623 \pm 0.092$} & {$38.250 \pm 0.071$} & {$74.029 \pm 0.128$} & {$57.161 \pm 0.084$} & {$35.734 \pm 0.065$}\\
\hline
\multicolumn{6}{c}{ }\\
\hline
\multicolumn{6}{|c|}{$3d$ ideal trees}\\
\hline
{3} & {$  0.709 \pm 0.003$} & {$ 0.823 \pm 0.001$} & {$ 1.291 \pm 0.003$} & {$ 1.161 \pm 0.001$} & {$ 0.097 \pm 0.001$}\\
{5} & {$  1.450 \pm 0.015$} & {$ 1.220 \pm 0.001$} & {$ 2.030 \pm 0.002$} & {$ 1.692 \pm 0.001$} & {$ 0.464 \pm 0.001$}\\
{10} & {$  3.320 \pm 0.012$} & {$ 1.938 \pm 0.003$} & {$ 3.357 \pm 0.009$} & {$ 2.769 \pm 0.003$} & {$ 1.098 \pm 0.003$}\\
{20} & {$  7.305 \pm 0.006$} & {$ 3.036 \pm 0.002$} & {$ 5.393 \pm 0.004$} & {$ 4.399 \pm 0.002$} & {$ 2.131 \pm 0.002$}\\
{30} & {$ 11.256 \pm 0.007$} & {$ 3.923 \pm 0.003$} & {$ 7.054 \pm 0.005$} & {$ 5.715 \pm 0.003$} & {$ 2.971 \pm 0.003$}\\
{45} & {$ 17.202 \pm 0.008$} & {$ 5.025 \pm 0.004$} & {$ 9.137 \pm 0.007$} & {$ 7.358 \pm 0.004$} & {$ 4.017 \pm 0.004$}\\
{75} & {$ 29.120 \pm 0.012$} & {$ 6.800 \pm 0.006$} & {$12.512 \pm 0.011$} & {$10.008 \pm 0.006$} & {$ 5.706 \pm 0.006$}\\
{150} & {$ 58.874 \pm 0.017$} & {$10.124 \pm 0.010$} & {$18.897 \pm 0.021$} & {$14.974 \pm 0.010$} & {$ 8.874 \pm 0.011$}\\
{230} & {$ 90.614 \pm 0.019$} & {$12.834 \pm 0.012$} & {$24.137 \pm 0.023$} & {$19.027 \pm 0.013$} & {$11.453 \pm 0.011$}\\
{450} & {$177.882 \pm 0.026$} & {$18.510 \pm 0.018$} & {$35.180 \pm 0.037$} & {$27.526 \pm 0.020$} & {$16.879 \pm 0.019$}\\
{900} & {$356.425 \pm 0.051$} & {$26.740 \pm 0.038$} & {$51.330 \pm 0.069$} & {$39.850 \pm 0.043$} & {$24.724 \pm 0.039$}\\
{1800} & {$713.491 \pm 0.101$} & {$38.243 \pm 0.079$} & {$74.043 \pm 0.129$} & {$57.174 \pm 0.090$} & {$35.723 \pm 0.073$}\\
\hline
\end{tabular}
\caption{
\label{tab:ConnectivityDataIdealTrees}
Connectivity and branching statistics of $2d$ and $3d$ ideal lattice trees of mass $N$:
$\langle n_3 \rangle$, average number of three-functional nodes.
$\langle \delta L_{\mathrm{center}} \rangle$, average path distance from the central node.
$\langle \delta L_{\mathrm{center}}^{\mathrm{max}} \rangle$, average longest path distance from the central node.
$\langle L \rangle$, average path distance between nodes.
$\langle N_{br} \rangle $, average branch weight.
Notice that statistics of ideal trees are independent of space dimensionality.
}
\end{table*}

\begin{table*}
\begin{tabular}{|c|c|c|c|c|c|}
\hline
{$N$} & {$\langle n_3\rangle$} & {$\langle \delta L_{\mathrm{center}} \rangle$} & {$\langle \delta L_{\mathrm{center}}^{\mathrm{max}} \rangle$} & {$\langle L \rangle$} & {$\langle N_{br} \rangle $}\\
\hline
\multicolumn{6}{c}{ }\\
\hline
\multicolumn{6}{|c|}{$2d$ melt of trees, $\rho_K l_K^2 = 2$}\\
\hline
{3} & {$  0.682 \pm 0.004$} & {$ 0.829 \pm 0.001$} & {$ 1.318 \pm 0.004$} & {$ 1.165 \pm 0.001$} & {$ 0.106 \pm 0.001$}\\
{5} & {$  1.414 \pm 0.005$} & {$ 1.220 \pm 0.001$} & {$ 2.039 \pm 0.002$} & {$ 1.694 \pm 0.001$} & {$ 0.464 \pm 0.001$}\\
{10} & {$  3.348 \pm 0.007$} & {$ 1.935 \pm 0.002$} & {$ 3.333 \pm 0.004$} & {$ 2.764 \pm 0.001$} & {$ 1.096 \pm 0.002$}\\
{20} & {$  7.395 \pm 0.010$} & {$ 3.051 \pm 0.003$} & {$ 5.432 \pm 0.007$} & {$ 4.408 \pm 0.003$} & {$ 2.147 \pm 0.003$}\\
{30} & {$ 11.439 \pm 0.012$} & {$ 3.956 \pm 0.004$} & {$ 7.122 \pm 0.008$} & {$ 5.744 \pm 0.004$} & {$ 3.008 \pm 0.004$}\\
{45} & {$ 17.472 \pm 0.008$} & {$ 5.113 \pm 0.003$} & {$ 9.307 \pm 0.007$} & {$ 7.449 \pm 0.004$} & {$ 4.112 \pm 0.003$}\\
{75} & {$ 29.554 \pm 0.009$} & {$ 7.043 \pm 0.005$} & {$12.973 \pm 0.009$} & {$10.285 \pm 0.005$} & {$ 5.959 \pm 0.005$}\\
{150} & {$ 59.770 \pm 0.016$} & {$10.789 \pm 0.009$} & {$20.152 \pm 0.016$} & {$15.788 \pm 0.010$} & {$ 9.554 \pm 0.008$}\\
{230} & {$ 91.950 \pm 0.020$} & {$13.996 \pm 0.012$} & {$26.319 \pm 0.022$} & {$20.499 \pm 0.014$} & {$12.646 \pm 0.012$}\\
{450} & {$180.556 \pm 0.030$} & {$21.130 \pm 0.017$} & {$40.094 \pm 0.032$} & {$30.937 \pm 0.018$} & {$19.524 \pm 0.016$}\\
{900} & {$361.877 \pm 0.064$} & {$32.303 \pm 0.051$} & {$61.678 \pm 0.095$} & {$47.284 \pm 0.055$} & {$30.321 \pm 0.051$}\\
\hline
\multicolumn{6}{c}{ }\\
\hline
\multicolumn{6}{|c|}{$3d$ melt of trees, $\rho_K l_K^3 = 2$}\\
\hline
{3} & {$ 0.671 \pm 0.003$} & {$ 0.832 \pm 0.001$} & {$ 1.329 \pm 0.003$} & {$ 1.166 \pm 0.001$} & {$ 0.110 \pm 0.001$}\\
{5} & {$  1.411 \pm 0.004$} & {$ 1.220 \pm 0.001$} & {$ 2.035 \pm 0.001$} & {$ 1.694 \pm 0.001$} & {$ 0.464 \pm 0.001$}\\
{10} & {$  3.340 \pm 0.004$} & {$ 1.934 \pm 0.001$} & {$ 3.334 \pm 0.003$} & {$ 2.765 \pm 0.001$} & {$ 1.096 \pm 0.001$}\\
{20} & {$  7.368 \pm 0.006$} & {$ 3.040 \pm 0.002$} & {$ 5.404 \pm 0.004$} & {$ 4.401 \pm 0.002$} & {$ 2.137 \pm 0.002$}\\
{30} & {$ 11.409 \pm 0.007$} & {$ 3.930 \pm 0.002$} & {$ 7.065 \pm 0.005$} & {$ 5.719 \pm 0.002$} & {$ 2.980 \pm 0.002$}\\
{45} & {$ 17.425 \pm 0.009$} & {$ 5.067 \pm 0.004$} & {$ 9.208 \pm 0.008$} & {$ 7.399 \pm 0.004$} & {$ 4.062 \pm 0.004$}\\
{75} & {$ 29.510 \pm 0.012$} & {$ 6.884 \pm 0.006$} & {$12.667 \pm 0.011$} & {$10.099 \pm 0.006$} & {$ 5.789 \pm 0.006$}\\
{150} & {$ 59.650 \pm 0.019$} & {$10.361 \pm 0.009$} & {$19.334 \pm 0.017$} & {$15.263 \pm 0.010$} & {$ 9.110 \pm 0.009$}\\
{230} & {$ 91.770 \pm 0.019$} & {$13.245 \pm 0.012$} & {$24.894 \pm 0.023$} & {$19.554 \pm 0.015$} & {$11.866 \pm 0.012$}\\
{450} & {$180.254 \pm 0.031$} & {$19.282 \pm 0.019$} & {$36.630 \pm 0.034$} & {$28.564 \pm 0.022$} & {$17.642 \pm 0.019$}\\
{900} & {$361.189 \pm 0.077$} & {$28.314 \pm 0.053$} & {$54.216 \pm 0.097$} & {$41.996 \pm 0.058$} & {$26.288 \pm 0.054$}\\
\hline
\multicolumn{6}{c}{ }\\
\hline
\multicolumn{6}{|c|}{$3d$ melt of trees, $\rho_K l_K^3 = 5$}\\
\hline
{3} & {$  0.672 \pm 0.004$} & {$ 0.832 \pm 0.001$} & {$ 1.329 \pm 0.004$} & {$ 1.166 \pm 0.001$} & {$ 0.110 \pm 0.001$}\\
{5} & {$  1.367 \pm 0.025$} & {$ 1.227 \pm 0.001$} & {$ 2.047 \pm 0.003$} & {$ 1.701 \pm 0.001$} & {$ 0.473 \pm 0.002$}\\
{10} & {$  3.285 \pm 0.012$} & {$ 1.942 \pm 0.003$} & {$ 3.349 \pm 0.009$} & {$ 2.774 \pm 0.003$} & {$ 1.103 \pm 0.003$}\\
{20} & {$  7.209 \pm 0.006$} & {$ 3.053 \pm 0.002$} & {$ 5.434 \pm 0.004$} & {$ 4.418 \pm 0.002$} & {$ 2.147 \pm 0.002$}\\
{30} & {$ 11.167 \pm 0.006$} & {$ 3.940 \pm 0.003$} & {$ 7.098 \pm 0.006$} & {$ 5.737 \pm 0.003$} & {$ 2.988 \pm 0.003$}\\
{45} & {$ 17.093 \pm 0.008$} & {$ 5.051 \pm 0.004$} & {$ 9.192 \pm 0.009$} & {$ 7.392 \pm 0.004$} & {$ 4.042 \pm 0.004$}\\
{75} & {$ 28.949 \pm 0.011$} & {$ 6.861 \pm 0.006$} & {$12.629 \pm 0.010$} & {$10.082 \pm 0.006$} & {$ 5.765 \pm 0.005$}\\
{150} & {$ 58.610 \pm 0.015$} & {$10.242 \pm 0.009$} & {$19.127 \pm 0.017$} & {$15.125 \pm 0.010$} & {$ 8.986 \pm 0.010$}\\
{230} & {$ 90.164 \pm 0.020$} & {$13.034 \pm 0.012$} & {$24.514 \pm 0.020$} & {$19.289 \pm 0.013$} & {$11.655 \pm 0.010$}\\
{450} & {$177.134 \pm 0.027$} & {$18.955 \pm 0.019$} & {$36.011 \pm 0.034$} & {$28.104 \pm 0.021$} & {$17.309 \pm 0.018$}\\
{900} & {$354.864 \pm 0.062$} & {$27.738 \pm 0.040$} & {$53.183 \pm 0.064$} & {$41.209 \pm 0.045$} & {$25.716 \pm 0.034$}\\
{1800} & {$710.024 \pm 0.110$} & {$39.850 \pm 0.082$} & {$78.622 \pm 0.154$} & {$59.424 \pm 0.092$} & {$37.310 \pm 0.077$}\\
\hline
\end{tabular}
\caption{
\label{tab:ConnectivityDataMeltTrees}
Connectivity and branching statistics of $2d$ and $3d$ melts of trees of mass $N$ with average density of Kuhn segments per unit cell $\rho_K l_K^d$.
Symbols as in Table~S\ref{tab:ConnectivityDataIdealTrees}.
}
\end{table*}

\begin{table*}
\begin{tabular}{|c|c|c|c|c|}
\hline
{} & $\langle \delta L_{\mathrm{center}} \rangle \sim N^{\rho}$ & $\langle \delta L_{\mathrm{center}}^{\mathrm{max}} \rangle\sim N^{\rho}$ & $\langle L \rangle \sim N^{\rho}$ & $\langle N_{br} \rangle \sim N^{\epsilon}$\\
\hline
\multicolumn{5}{c}{}\\
\hline
\multicolumn{5}{|c|}{Ideal trees}\\
\hline
$\Delta$ & $0.239 \pm 0.045$ & $0.280 \pm 0.044$ & $0.271 \pm 0.028$ & $0.706 \pm 0.032$\\
${\tilde \chi}^2$ & $1.672$ & $1.073$ & $2.355$ & $2.706$\\
${\mathcal Q}$ & $0.123$ & $0.376$ & $0.028$ & $0.013$\\
& $\rho = 0.446 \pm 0.013$ & $\rho = 0.467 \pm 0.010$ & $\rho = 0.457 \pm 0.007$ & $\epsilon = 0.529 \pm 0.002$\\
\hline
$\Delta$ & $0$ & $0$ & $0$ & $0$\\
${\tilde \chi}^2$ & $7.424$ & $11.399$ & $10.476$ & $12.172$\\
${\mathcal Q}$ & $0.006$ & $0.001$ & $0.001$ & $0.001$\\
& $\rho = 0.525 \pm 0.002$ & $\rho = 0.538 \pm 0.001$ & $\rho = 0.529 \pm 0.001$ & $\epsilon = 0.543 \pm 0.002$\\
\hline
& \multicolumn{3}{c|}{$\mathbf{\rho = 0.494 \pm 0.013 \pm 0.038}$} & $\mathbf{\epsilon = 0.536 \pm 0.002 \pm 0.007}$\\
\hline
\multicolumn{5}{c}{}\\
\hline
\multicolumn{5}{|c|}{$2d$ melt of trees, $\rho_K l_K^2 = 2$}\\
\hline
$\Delta$ & $0.767 \pm 0.045$ & $0.726 \pm 0.048$ & $0.696 \pm 0.028$ & $0.871 \pm 0.069$\\
${\tilde \chi}^2$ & $2.770$ & $1.612$ & $4.861$ & $0.807$\\
${\mathcal Q}$ & $0.017$ & $0.153$ & $0.0002$ & $0.544$\\
& $\rho = 0.606 \pm 0.002$ & $\rho = 0.616 \pm 0.002$ & $0.603 \pm 0.002$ & $\epsilon = 0.623 \pm 0.002$\\
\hline
$\Delta$ & $0$ & $0$ & $0$ & $0$\\
${\tilde \chi}^2$ & $0.153$ & $2.815$ & $0.190$ & $10.607$\\
${\mathcal Q}$ & $0.696$ & $0.093$ & $0.663$ & $0.001$\\
& $\rho = 0.613 \pm 0.001$ & $\rho = 0.625 \pm 0.001$ & $\rho = 0.613 \pm 0.001$ & $\epsilon = 0.643 \pm 0.001$\\
\hline
& \multicolumn{3}{c|}{$\mathbf{\rho = 0.613 \pm 0.002 \pm 0.007}$} & $\mathbf{\epsilon = 0.633 \pm 0.002 \pm 0.010}$\\
\hline
\multicolumn{5}{c}{}\\
\hline
\multicolumn{5}{|c|}{$3d$ melt of trees, $\rho_K l_K^3 = 2$}\\
\hline
$\Delta$ & $0.192 \pm 0.061$ & $0.289 \pm 0.046$ & $0.248 \pm 0.033$ & $0.693 \pm 0.048$\\
${\tilde \chi}^2$ & $3.046$ & $1.795$ & $3.303$ & $3.963$\\
${\mathcal Q}$ & $0.009$ & $0.110$ & $0.006$ & $0.001$\\
& $\rho = 0.450 \pm 0.023$ & $\rho = 0.497 \pm 0.012$ & $0.477 \pm 0.010$ & $\epsilon = 0.548 \pm 0.003$\\
\hline
$\Delta$ & $0$ & $0$ & $0$ & $0$\\
${\tilde \chi}^2$ & $1.646$ & $6.011$ & $7.073$ & $11.762$\\
${\mathcal Q}$ & $0.200$ & $0.014$ & $0.008$ & $0.001$\\
& $\rho = 0.558 \pm 0.001$ & $\rho = 0.572 \pm 0.001$ & $\rho = 0.561 \pm 0.001$ & $\epsilon = 0.586 \pm 0.002$\\
\hline
& \multicolumn{3}{c|}{$\mathbf{\rho = 0.519 \pm 0.023 \pm 0.047}$} & $\mathbf{\epsilon = 0.567 \pm 0.003 \pm 0.019}$\\
\hline
\multicolumn{5}{c}{}\\
\hline
\multicolumn{5}{|c|}{$3d$ melt of trees, $\rho_K l_K^3 = 5$}\\
\hline
$\Delta$ & $0.301 \pm 0.037$ & $0.499 \pm 0.036$ & $0.344 \pm 0.023$ & $0.737 \pm 0.031$\\
${\tilde \chi}^2$ & $2.724$ & $1.534$ & $3.381$ & $0.893$\\
${\mathcal Q}$ & $0.012$ & $0.163$ & $0.003$ & $0.499$\\
& $\rho = 0.493 \pm 0.008$ & $\rho = 0.540 \pm 0.003$ & $0.500 \pm 0.004$ & $\epsilon = 0.548 \pm 0.002$\\
\hline
$\Delta$ & $0$ & $0$ & $0$ & $0$\\
${\tilde \chi}^2$ & $25.137$ & $0.095$ & $36.149$ & $45.573$\\
${\mathcal Q}$ & $10^{-6}$ & $0.758$ & $<10^{-6}$ & $<10^{-6}$\\
& $\rho = 0.539 \pm 0.002$ & $\rho = 0.563 \pm 0.001$ & $\rho = 0.543 \pm 0.001$ & $\epsilon = 0.559 \pm 0.002$\\
\hline
& \multicolumn{3}{c|}{$\mathbf{\rho = 0.530 \pm 0.008 \pm 0.025}$} & $\mathbf{\epsilon = 0.554 \pm 0.002 \pm 0.006}$\\
\hline
\end{tabular}
\caption{
\label{tab:FitsCritExpsTreeStructure}
Critical exponents $\rho$ and $\epsilon$,
describing path length ($\langle \delta L_{\mathrm{center}} \rangle \sim \langle \delta L_{\mathrm{center}}^{\mathrm{max}} \rangle \sim \langle L \rangle \sim N^{\rho}$)
and branching statistics ($\langle N_{br}(N) \rangle \sim N^{\epsilon}$), respectively.
Single estimates were obtained from best fits of
data with $N \geq 230$ ($2d$ and $3d$ melt of trees for $\rho_K l_K^d = 2$) and $N \geq 450$ (ideal trees, $3d$ melt of trees for $\rho_K l_K^3 = 5$)
to simple power-law behaviour ($\Delta=0$) and
data with $N \geq 10$ to power-law behaviour with a correction-to-scaling term ($\Delta > 0$).
Final estimates (in boldface) show the corresponding averages.
Uncertainties are reported as ``$\pm \mbox{ statistical error} \pm \mbox{systematic error}$''.
For more details on fitting procedures and error estimation, see Sec.~IIIC 
main paper.
}
\end{table*}

\begin{table*}
\begin{tabular}{|c|c|c|c|c|}
\hline
{$N$} & {$\langle R^2 (\ell = {\tt nint} (\langle L(N) \rangle)) \rangle$} & {$\langle L_{\mathrm{max}} \rangle$} & {$\langle R^2 (L_{\mathrm{max}}) \rangle$} & {$\left \langle R_g^2 \right \rangle $} \\
\hline
\multicolumn{5}{c}{ }\\
\hline
\multicolumn{5}{|c|}{$2d$ ideal trees}\\
\hline
{3} & {$1.000 \pm 0.000$} & {$2.282 \pm 0.004$} & {$2.277 \pm 0.016$} & {$ 0.581 \pm 0.002$} \\
{5} & {$2.003 \pm 0.006$} & {$3.571 \pm 0.005$} & {$3.518 \pm 0.029$} & {$ 0.845 \pm 0.003$} \\
{10} & {$3.005 \pm 0.009$} & {$6.225 \pm 0.007$} & {$6.300 \pm 0.054$} & {$ 1.389 \pm 0.005$} \\
{20} & {$3.999 \pm 0.011$} & {$10.290 \pm 0.012$} & {$10.100 \pm 0.088$} & {$ 2.195 \pm 0.008$} \\
{30} & {$5.986 \pm 0.019$} & {$13.617 \pm 0.016$} & {$13.493 \pm 0.115$} & {$ 2.851 \pm 0.011$} \\
{45} & {$7.017 \pm 0.013$} & {$17.755 \pm 0.014$} & {$17.666 \pm 0.096$} & {$ 3.684 \pm 0.009$} \\
{75} & {$10.018 \pm 0.018$} & {$24.545 \pm 0.018$} & {$24.695 \pm 0.126$} & {$ 5.031 \pm 0.011$} \\
{150} & {$15.008 \pm 0.031$} & {$37.266 \pm 0.032$} & {$37.262 \pm 0.215$} & {$ 7.500 \pm 0.022$} \\
{230} & {$18.972 \pm 0.038$} & {$47.726 \pm 0.040$} & {$47.515 \pm 0.267$} & {$ 9.467 \pm 0.022$} \\
{450} & {$26.956 \pm 0.054$} & {$69.778 \pm 0.060$} & {$70.048 \pm 0.398$} & {$13.720 \pm 0.034$} \\
{900} & {$39.990 \pm 0.113$} & {$102.128 \pm 0.125$} & {$102.987 \pm 0.831$} & {$19.904 \pm 0.064$} \\
{1800} & {$57.094 \pm 0.227$} & {$147.554 \pm 0.257$} & {$148.062 \pm 1.680$} & {$28.630 \pm 0.145$} \\
\hline
\multicolumn{5}{c}{ }\\
\hline
\multicolumn{5}{|c|}{$3d$ ideal trees}\\
\hline
{3} & {$1.000 \pm 0.000$} &  {$2.291 \pm 0.004$} & {$2.276 \pm 0.012$} & {$ 0.578 \pm 0.001$} \\
{5} & {$1.994 \pm 0.006$} & {$3.566 \pm 0.007$} & {$3.617 \pm 0.033$} & {$ 0.851 \pm 0.010$} \\
{10} & {$2.978 \pm 0.015$} & {$6.257 \pm 0.014$} & {$6.207 \pm 0.085$} & {$ 1.323 \pm 0.028$} \\
{20} & {$3.998 \pm 0.006$} & {$10.287 \pm 0.009$} & {$10.305 \pm 0.052$} & {$ 2.199 \pm 0.005$} \\
{30} & {$5.998 \pm 0.011$} & {$13.615 \pm 0.011$} & {$13.604 \pm 0.069$} & {$ 2.857 \pm 0.006$} \\
{45} & {$7.018 \pm 0.012$} & {$17.774 \pm 0.015$} & {$17.668 \pm 0.090$} & {$ 3.688 \pm 0.008$} \\
{75} & {$10.012 \pm 0.018$} & {$24.526 \pm 0.022$} & {$24.587 \pm 0.126$} & {$ 5.006 \pm 0.012$} \\
{150} & {$14.974 \pm 0.027$} & {$37.291 \pm 0.035$} & {$37.094 \pm 0.192$} & {$ 7.468 \pm 0.016$} \\
{230} & {$19.034 \pm 0.035$} & {$47.774 \pm 0.045$} & {$47.664 \pm 0.250$} & {$ 9.528 \pm 0.021$} \\
{450} & {$27.931 \pm 0.052$} & {$69.862 \pm 0.067$} & {$69.747 \pm 0.362$} & {$13.718 \pm 0.028$} \\
{900} & {$39.923 \pm 0.104$} & {$102.155 \pm 0.139$} & {$103.217 \pm 0.764$} & {$19.936 \pm 0.063$} \\
{1800} & {$57.296 \pm 0.208$} & {$147.584 \pm 0.283$} & {$149.921 \pm 1.575$} & {$28.802 \pm 0.122$} \\
\hline
\end{tabular}
\caption{
\label{tab:ConformationalDataIdealTrees}
Conformational statistics of $2d$ and $3d$ ideal lattice trees of mass $N$:
$\langle R^2 ( \ell = {\tt nint}( \langle L(N) \rangle)) \rangle$, average-square end-to-end distance of paths of length $\ell = {\tt nint}(\langle L(N) \rangle) \equiv$ closest-integer-to $\langle L(N) \rangle$.
$\langle L_{\mathrm{max}} \rangle$, average length of the longest paths.
$\langle R^2 (L_{\mathrm{max}}) \rangle$, average-square end-to-end distance of the longest paths.
$\left \langle R_g^2 \right \rangle$, average-square gyration radius.
As in Table~S\ref{tab:ConnectivityDataIdealTrees},
statistics of ideal trees is independent of space dimensionality.  
}
\end{table*}

\begin{table*}
\begin{tabular}{|c|c|c|c|c|}
\hline
{$N$} & {$\langle R^2 (\ell = {\tt nint} (\langle L(N) \rangle)) \rangle$} & {$\langle L_{\mathrm{max}} \rangle$} & {$\langle R^2 (L_{\mathrm{max}}) \rangle$} & {$\left \langle R_g^2 \right \rangle $} \\
\hline
\multicolumn{5}{c}{ }\\
\hline
\multicolumn{5}{|c|}{$2d$ melt of trees, $\rho_K l_K^2 = 2$}\\
\hline
{3} & {$1.000 \pm 0.000$} & {$2.318 \pm 0.004$} & {$2.453 \pm 0.017$} & {$0.603 \pm 0.002$} \\
{5} & {$2.181 \pm 0.005$} & {$3.587 \pm 0.005$} & {$4.102 \pm 0.031$} & {$0.933 \pm 0.003$} \\
{10} & {$3.690 \pm 0.008$} & {$6.2 38 \pm 0.007$} & {$8.250 \pm 0.061$} & {$1.730 \pm 0.005$} \\
{20} & {$5.552 \pm 0.009$} & {$10.3 53 \pm 0.012$} & {$16.281 \pm 0.122$} & {$3.237 \pm 0.009$} \\
{30} & {$9.597 \pm 0.018$} & {$13.7 42 \pm 0.016$} & {$24.101 \pm 0.172$} & {$4.677 \pm 0.015$} \\
{45} & {$12.083 \pm 0.012$} & {$18.115 \pm 0.014$} & {$35.009 \pm 0.157$} & {$6.733 \pm 0.012$} \\
{75} & {$20.219 \pm 0.019$} & {$25.4 49 \pm 0.018$} & {$57.527 \pm 0.233$} & {$10.794 \pm 0.014$} \\
{150} & {$40.727 \pm 0.043$} & {$39.801 \pm 0.033$} & {$112.627 \pm 0.508$} & {$20.585 \pm 0.031$} \\
{230} & {$57.661 \pm 0.059$} & {$52.140 \pm 0.045$} & {$170.895 \pm 0.778$} & {$30.759 \pm 0.052$} \\
{450} & {$113.568 \pm 0.111$} & {$79.688 \pm 0.064$} & {$328.627 \pm 1.392$} & {$58.535 \pm 0.074$} \\
{900} & {$220.202 \pm 0.396$} & {$122.854 \pm 0.189$} & {$651.115 \pm 5.217$} & {$115.109 \pm 0.293$} \\
\hline
\multicolumn{5}{c}{ }\\
\hline
\multicolumn{5}{|c|}{$3d$ melt of trees, $\rho_K l_K^3 = 2$}\\
\hline
{3} & {$1.000 \pm 0.000$} & {$2.329 \pm 0.003$} & {$2.404 \pm 0.010$} & {$0.596 \pm 0.001$} \\
{5} & {$2.099 \pm 0.003$} & {$3.589 \pm 0.004$} & {$3.907 \pm 0.017$} & {$0.897 \pm 0.002$} \\
{10} & {$3.406 \pm 0.005$} & {$6.233 \pm 0.006$} & {$7.423 \pm 0.037$} & {$1.583 \pm 0.003$} \\
{20} & {$4.849 \pm 0.005$} & {$10.313 \pm 0.008$} & {$13.348 \pm 0.056$} & {$2.736 \pm 0.005$} \\
{30} & {$7.773 \pm 0.009$} & {$13.625 \pm 0.010$} & {$18.401 \pm 0.073$} & {$3.722 \pm 0.006$} \\
{45} & {$9.373 \pm 0.012$} & {$17.918 \pm 0.016$} & {$25.342 \pm 0.117$} & {$5.037 \pm 0.009$} \\
{75} & {$14.213 \pm 0.019$} & {$24.831 \pm 0.022$} & {$36.610 \pm 0.170$} & {$7.233 \pm 0.012$} \\
{150} & {$22.944 \pm 0.029$} & {$38.172 \pm 0.033$} & {$60.550 \pm 0.270$} & {$11.751 \pm 0.024$} \\
{230} & {$32.153 \pm 0.046$} & {$49.289 \pm 0.045$} & {$80.884 \pm 0.383$} & {$15.732 \pm 0.035$} \\
{450} & {$50.246 \pm 0.072$} & {$72.765 \pm 0.069$} & {$128.880 \pm 0.616$} & {$24.744 \pm 0.047$} \\
{900} & {$77.753 \pm 0.205$} & {$107.929 \pm 0.193$} & {$202.464 \pm 1.867$} & {$38.813 \pm 0.166$} \\
\hline
\multicolumn{5}{c}{ }\\
\hline
\multicolumn{5}{|c|}{$3d$ melt of trees, $\rho_K l_K^3 = 5$}\\
\hline
{3} & {$1.000 \pm 0.000$} & {$2.329 \pm 0.004$} & {$2.385 \pm 0.013$} & {$0.591 \pm 0.001$} \\
{5} & {$2.031 \pm 0.006$} & {$3.628 \pm 0.007$} & {$3.710 \pm 0.033$} & {$0.868 \pm 0.012$} \\
{10} & {$3.161 \pm 0.014$} & {$6.260 \pm 0.014$} & {$6.789 \pm 0.088$} & {$1.454 \pm 0.017$} \\
{20} & {$4.305 \pm 0.006$} & {$10.372 \pm 0.009$} & {$11.488 \pm 0.055$} & {$2.406 \pm 0.005$} \\
{30} & {$6.636 \pm 0.011$} & {$13.697 \pm 0.012$} & {$15.403 \pm 0.075$} & {$3.181 \pm 0.006$} \\
{45} & {$7.875 \pm 0.012$} & {$17.885 \pm 0.016$} & {$20.719 \pm 0.100$} & {$4.191 \pm 0.008$} \\
{75} & {$11.572 \pm 0.018$} & {$24.757 \pm 0.022$} & {$29.073 \pm 0.143$} & {$5.849 \pm 0.012$} \\
{150} & {$18.007 \pm 0.029$} & {$37.753 \pm 0.035$} & {$46.095 \pm 0.227$} & {$9.110 \pm 0.017$} \\
{230} & {$23.296 \pm 0.037$} & {$48.534 \pm 0.045$} & {$60.730 \pm 0.300$} & {$11.871 \pm 0.024$} \\
{450} & {$35.781 \pm 0.057$} & {$71.519 \pm 0.068$} & {$91.625 \pm 0.456$} & {$17.953 \pm 0.035$} \\
{900} & {$54.530 \pm 0.121$} & {$105.872 \pm 0.147$} & {$141.131 \pm 0.990$} & {$27.344 \pm 0.077$} \\
{1800} & {$83.166 \pm 0.257$} & {$153.467 \pm 0.296$} & {$219.536 \pm 2.196$} & {$41.237 \pm 0.164$} \\
\hline
\end{tabular}
\caption{
\label{tab:ConformationalDataMeltTrees}
Conformational statistics of $2d$ and $3d$ melts of trees of mass $N$ with average density of Kuhn segments per unit cell $\rho_K l_K^d$.
Symbols as in Table~S\ref{tab:ConformationalDataIdealTrees}.}
\end{table*}

\begin{table*}
\begin{tabular}{|c|c|c|c|}
\hline
{} & $\langle R^2 (\ell={\tt nint} (\langle L(N) \rangle)) \rangle \sim \ell^{2\nu_{\mathrm{path}}}$ & $\langle R^2 (L_{\mathrm{max}}) \rangle \sim \langle L_{\mathrm{max}} \rangle^{2\nu_{\mathrm{path}}}$ & $\left \langle R_g^2 \right \rangle \sim N^{2\nu}$ \\
\hline
\multicolumn{4}{c}{ }\\
\hline
\multicolumn{4}{|c|}{Ideal trees}\\
\hline
$\Delta$ & -- & -- & $0.277 \pm 0.105$ \\
${\tilde \chi}^2$ & -- & -- & $1.770$ \\
${\mathcal Q}$ & -- & -- & $0.101$ \\
& -- & -- & $\nu = 0.229 \pm 0.013$ \\
\hline
$\Delta$ & $0$ & $0$ & $0$ \\
${\tilde \chi}^2$ & $0.999$ & $0.112$ & $0.562$ \\
${\mathcal Q}$ & $0.318$ & $0.738$ & $0.454$ \\
& $\nu_{\mathrm{path}} = 0.504 \pm 0.003$ & $\nu_{\mathrm{path}} = 0.513 \pm 0.007$ & $\nu = 0.268 \pm 0.002$ \\
\hline
& \multicolumn{2}{|c|}{$\mathbf{\nu_{\mathrm{path}} = 0.509 \pm 007 \pm 0.005}$} & $\mathbf{\nu = 0.249 \pm 0.013 \pm 0.020}$ \\
\hline
\multicolumn{4}{c}{ }\\
\hline
\multicolumn{4}{|c|}{$2d$ melt of trees, $\rho_K l_K^2 = 2$}\\
\hline
$\Delta$ & -- & -- & -- \\
${\tilde \chi}^2$ & -- & -- & -- \\
${\mathcal Q}$ & -- & -- & -- \\
& -- & -- & -- \\
\hline
$\Delta$ & $0$ & $0$ & $0$ \\
${\tilde \chi}^2$ & $44.038$ & $1.709$ & $8.515$ \\
${\mathcal Q}$ & $< 10^{-6}$ & $0.191$ & $0.004$ \\
& $\nu_{\mathrm{path}} = 0.781 \pm 0.001$ & $\nu_{\mathrm{path}} = 0.778 \pm 0.005$ & $\nu = 0.483 \pm 0.001$ \\
\hline
& \multicolumn{2}{|c|}{$\mathbf{\nu_{\mathrm{path}} = 0.780 \pm 0.005 \pm 0.002}$} & $\mathbf{\nu = 0.483 \pm (0.008) \pm (0.013)}$ \\
\hline
\multicolumn{4}{c}{ }\\
\hline
\multicolumn{4}{|c|}{$3d$ melt of trees, $\rho_K l_K^3 = 2$}\\
\hline
$\Delta$ & -- & -- & $0.384 \pm 0.086$ \\
${\tilde \chi}^2$ & -- & -- & $1.469$ \\
${\mathcal Q}$ & -- & -- & $0.196$ \\
& -- & -- & $\nu = 0.309 \pm 0.008$ \\
\hline
$\Delta$ & $0$ & $0$ & $0$ \\
${\tilde \chi}^2$ & $4.087$ & $1.951$ & $6.973$ \\
${\mathcal Q}$ & $0.043$ & $0.163$ & $0.001$ \\
& $\nu_{\mathrm{path}} = 0.597 \pm 0.002$ & $\nu_{\mathrm{path}} = 0.589 \pm 0.006$ & $\nu = 0.336 \pm 0.001$ \\
\hline
& \multicolumn{2}{|c|}{$\mathbf{\nu_{\mathrm{path}} = 0.593 \pm 0.006 \pm 0.004}$} & $\mathbf{\nu = 0.323 \pm 0.008 \pm 0.013}$ \\
\hline
\multicolumn{4}{c}{ }\\
\hline
\multicolumn{4}{|c|}{$3d$ melt of trees, $\rho_K l_K^3 = 5$}\\
\hline
$\Delta$ & -- & -- & $0.391 \pm 0.082$ \\
${\tilde \chi}^2$ & -- & -- & $0.508$ \\
${\mathcal Q}$ & -- & -- & $0.803$ \\
& -- & -- & $\nu = 0.286 \pm 0.007$ \\
\hline
$\Delta$ & $0$ & $0$ & $0$ \\
${\tilde \chi}^2$ & $13.089$ & $3.524$ & $1.927$ \\
${\mathcal Q}$ & $0.0003$ & $0.061$ & $0.165$ \\
& $\nu_{\mathrm{path}} = 0.563 \pm 0.002$ & $\nu_{\mathrm{path}} = 0.567 \pm 0.007$ & $\nu = 0.301 \pm 0.002$ \\
\hline
& \multicolumn{2}{|c|}{$\mathbf{\nu_{\mathrm{path}} = 0.565 \pm 0.007 \pm 0.002}$} & $\mathbf{\nu = 0.294 \pm 0.007 \pm 0.008}$ \\
\hline
\end{tabular}
\caption{
\label{tab:FitsCritExpsSpatialStructure}
Critical exponents $\nu_{\mathrm{path}}$ and $\nu$ 
describing, respectively, the scaling behaviors of
$\langle R^2(\ell={\tt nint}(\langle L(N) \rangle)) \rangle \sim \ell^{2\nu_{\mathrm{path}}}$ and
$\langle R^2(L_{\mathrm{max}}) \rangle \sim \langle L_{\mathrm{max}}\rangle^{2\nu_{\mathrm{path}}}$,
and
$\langle R_g^2(N) \rangle \sim \langle N\rangle^{2\nu}$.
Single estimates for $\nu$ are the results of fitting the data to
three- ($\Delta > 0$ for data with $N \geq 10$) and
two-parameter ($\Delta = 0$ for data with:
$N \geq 450$ (ideal trees, $3d$ melt of trees for $\rho_K l_K^3 = 5$) and $N \geq 230$ ($2d$ and $3d$ melt of trees for $\rho_K l_K^d = 2$))
functions, as described in Sec.~IIIC 
main paper.
In the case of $2d$ melts where the three-parameter fit fails,
corresponding statistical and systematic errors (in brackets) have been set to the same trends observed for $3d$ melt of trees for $\rho_K l_K^3 = 2$.
For the critical exponent $\nu_{\mathrm{path}}$,
we have attempted similar analysis with two- and three-parameter fit functions:
$\log \langle R^2(\ell={\tt nint}(\langle L(N) \rangle)) \rangle = a + 2\nu_{\mathrm{path}} \log l$ and
$\log \langle R^2(\ell={\tt nint}(\langle L(N) \rangle)) \rangle = a + b \, l^{-\Delta_0} + 2\nu_{\mathrm{path}} \log l - b (\Delta-\Delta_0) \, l^{-\Delta_0} \log l$,
and analogous expressions for $\langle L_{\mathrm{max}} \rangle$.
Since three-parameter fits fail to produce reliable results,
only estimates from two-parameter fits have been combined into our final values for $\nu_{\mathrm{path}}$. 
}
\end{table*}

\end{document}